\documentclass[a4paper,fleqn,usenatbib]{mnras}

\usepackage{newtxtext, newtxmath}
\usepackage{mathptmx}
\usepackage{txfonts}

\usepackage[T1]{fontenc}
\usepackage{ae,aecompl}
\usepackage{caption}
\usepackage{subcaption}
\usepackage{url}
\usepackage{float}
\usepackage{multirow}
\usepackage{lscape}
\usepackage[figuresright]{rotating}
\usepackage{graphicx}	

\title[ICL at the Frontier]{Intracluster Light at the Frontier II: The Frontier Fields Clusters}

\author[M. Montes \& I. Trujillo ]{
Mireia Montes$^{1}$\thanks{E-mail: mireia.montes.quiles@gmail.com (MM)}
 and Ignacio Trujillo$^{2,3}$
\\
$^{1}$Department of Astronomy, Yale University, 06511 New Haven, CT, USA\\
$^{2}$Instituto de Astrof\'{\i}sica de Canarias,c/ V\'{\i}a L\'actea s/n, E38205 - La Laguna, Tenerife, Spain\\
$^{3}$Departamento de Astrof\'isica, Universidad de La Laguna, E38205 La Laguna, Tenerife, Spain
}

\date{Accepted XXX. Received YYY; in original form ZZZ}

\pubyear{2017}

\begin{document}

\label{firstpage}
\pagerange{\pageref{firstpage}--\pageref{lastpage}}
\maketitle


\begin{abstract}

Multiwavelength deep observations are a key tool to understand the origin of the diffuse light in clusters of galaxies: the intra-cluster light (ICL). For this reason, we take advantage of the Hubble Frontier Fields survey to investigate the properties of the stellar populations of the ICL of its $6$ massive intermediate redshift (0.3$<$z$<$0.6) clusters. We carry on this analysis down to a radial distance of $\sim 120$ kpc from the brightest cluster galaxy. We found that the average metallicity of the ICL is [Fe/H]$_{ICL} \sim -0.5$, compatible with the value of the outskirts of the Milky Way. The mean stellar ages of the ICL are between $2$ to $6$ Gyr younger than the most massive galaxies of the clusters. Those results suggest that the ICL of these massive ($> 10^{15}\, M_{\odot}$) clusters  is formed by the stripping of MW-like objects that have been accreted at $z<1$, in agreement with current simulations. We do not find any significant increase in the fraction of light of the ICL with cosmic time, although the redshift range explored is narrow to derive any strong conclusion. When exploring the slope of the stellar mass density profile, we found that the ICL of the HFF clusters follows the shape of their underlying dark matter haloes, in agreement with the idea that the ICL is the result of the stripping of galaxies at recent times. 
\end{abstract}




\section{Introduction}

The most revealing signature of galaxy cluster assembly is contained within a diffuse component  occupying the space between the galaxies in the clusters. This component is composed of a substantial fraction of stars ($5-20\%$ of the total light of the cluster, \citealt[][]{Krick2007}) . These stars constitute the so-called intra-cluster light (ICL, see \citealt{Mihos2016} for a review). This diffuse light is thought to form primarily by galaxies that interact and merge during the hierarchical accretion history of the cluster \citep[e.g.][]{Gregg1998, Mihos2005, Conroy2007, Presotto2014, Contini2014}. 

Despite its enormous importance for understanding clusters, the ICL is still mostly unexplored. This component is extremely challenging to probe due to its very low surface brightness \citep[$\mu_V\gtrsim 26$ mag/arcsec$^2$, e.g.][]{Mihos2005, Zibetti2005, Rudick2010}. In addition, the ICL is normally contaminated by foreground and background (in projection) galaxies. Moreover, the separation between the ICL and the outer regions of the brightest central galaxies is an ill-defined problem \citep[e.g.][]{Gonzalez2005, Krick2007, Rudick2011, Jimenez-Teja2016}.

In order to comprehend the process of galaxy cluster evolution, it is important to determine how and when the ICL formed. In this sense, a useful tool to determine  the properties of the ICL is the study of its stellar populations. In fact, the ages and metallicities of the ICL population reflect the properties of the progenitor galaxies from which its stars got stripped. For example, \citet[][]{Contini2014} predicted that the bulk of the ICL light is produced by the most massive (M$_*\sim 10^{10-11}$M$_\odot$) galaxies as they fall into the cluster core \citep[see also][]{Rudick2011, Cooper2013}. If this is the case, the ICL should exhibit a mean metallicity similar to the outer regions of these massive satellites. Additionally, the age of the ICL stellar populations should give us an upper limit on when the formation of the ICL took place. This is because we do not expect any star formation in the ICL component after their stars have been stripped from their progenitor objects. In this sense, knowing the age and metallicity of the ICL of the clusters allow us to infer how (and when) the assembly history of these clusters was, ranging from the shredding of dwarf galaxies \citep[][]{Purcell2007, Contini2014}, to violent mergers with the central galaxies of the cluster \citep[][]{Murante2007, Conroy2007}, or in situ formation \citep[][]{Puchwein2010}. 

In addition to the age and metallicity of the ICL, probing how the amount of stellar mass of this component has changed with redshift indicates the growth speed of the clusters. Particularly interesting is the $z\sim0.5$ epoch, a period of time crucial to understand galaxy cluster evolution as it is expected that they may have accreted as much as half their mass by then \citep[e.g.][]{deLucia2007}. Simulations \citep[e.g.][]{Rudick2011, Contini2014} show that there is a strong evolution in the fraction of light contained in the ICL with respect to the total light of the cluster since $z\sim 0.5$. 
However, given both the ambiguity in defining the ICL and the observational difficulties in characterizing it, studies have found inconsistent results in the correlation between the fraction of light in this component and redshift or mass of the cluster \citep[e.g.][]{Lin2004, Zibetti2005, Krick2007, Guennou2012, Giallongo2014, Presotto2014, Burke2015}. Finally, the spatial distribution of the stars belonging to the ICL might also encode information about the entire assembly history of the halo it belongs to. Recently, \citet[][]{Pillepich2017b} showed that the slope of the stellar density profile of the ICL can help us understand the underlying dark matter halos \citep[see also][]{Pillepich2014}. 

The Hubble Frontier Fields\footnote{http://www.stsci.edu/hst/campaigns/frontier-fields} (HFF) survey represents the largest investment of HST time for deep observations of galaxy clusters. This survey consists in observations of $6$ very massive galaxy clusters in the redshift range $0.3<z<0.6$. With its incomparable depth, those images represent an excellent opportunity to study the properties of the ICL. Ages and metallicities of the ICL can be studied in detail taking advantage of the inclusion of very deep near-infrared (NIR) data to break the age-metallicity degeneracy \citep[e.g.][]{Anders2004}. 
The goal of this paper is to explore this question in detail and characterize the age and metallicity of the ICL of massive clusters at large radial distances from the centre (R$>50$ kpc) with unprecedented accuracy. Different scenarios for the origin of the ICL result in different stellar population properties and provide insightful evidence for the formation of both ICL and cluster. In our pilot project \citep[][hereafter Paper I]{MT14}, we demonstrated that we can derive the age and metallicity properties as well as the stellar mass fraction of the ICL for the HFF cluster Abell 2744. We found that the ICL is ($\sim3$ Gyr) younger  and more metal-poor than the center of the most massive galaxies of the cluster ($Z\sim Z_{\odot}$, while $Z_{gal}\sim2Z_{\odot}$).
The methodology applied in this paper differs from the one conducted in Paper I, where we relied in restframe colours to infer the age and metallicity radial profiles. In this paper, we are using the complete information given by all the broadband filters, from F435W to F160W.

Throughout this work, we adopt a standard cosmological model with the following parameters: $H_0$=70 km s$^{-1}$ Mpc$^{-1}$, $\Omega_m$=$0.3$ and $\Omega_\Lambda$=$0.7$. All magnitudes are in the AB magnitude system.

\section{Data}\label{data}

Our work is based in the complete HST data of the six HFF clusters (ID13495, PI: J. Lotz and ID13386). The ACS images were taken in the following filters: F435W,  F606W and F814W. NIR observations  include imaging in four filters F105W, F125W, F140W, F160W. The data were directly retrieved from the archive\footnote{http://www.stsci.edu/hst/campaigns/frontier-fields/FF-Data}. The HFF team reduced the data for each cluster using a two-step process. First, the exposures were reduced following standard HST procedures both for the ACS and WFC3 data, quality inspected, geometrically corrected and combined. Those two latter steps were performed using Astrodrizzle\footnote{http://drizzlepac.stsci.edu/}, while the alignment of the images has been done with Tweakreg. Once this step is complete, the images are reprocessed using all the information from all the exposures. Specifically this step includes: recalibration and bad pixel/cosmic-ray rejection, "self-calibration" of the ACS data, improve WFC3 flagging of pixels affected for persistence due to bright sources and reprocess those images affected by time variable sky, and combine the images of both cameras to produce the deepest images of the clusters.
The ACS WFC "self-calibration" step identifies and removes the pixels affected by charge transfer efficiency (CTE) resulting in a smoother image and a narrower pixel noise distribution. Also, a few of the HFF observations in the IR exhibit a time-variable sky background signal due to time variable atmospheric emission. The exposures were corrected from this variable emission and included into the final mosaics.  More information about the processing of the HFF data can be found here\footnote{\url{https://archive.stsci.edu/pub/hlsp/frontier/abell2744/ images/hst/v1.0-epoch2/hlsp_frontier_hst_abell2744_v1.0-epoch2_readme.pdf}}.
For both cameras, flat fields are claimed to be accurate to better than $1\%$ across the detector. The mosaics we used consist on drizzled science images with pixel size $0\farcs06$. In the case of the WFC3, this pixel size is closer to one half of the original pixel. 

To conduct our goals  a detailed process to avoid biases and contaminations is required. First, we corrected by the effect of the PSF across the different bands. It has been shown in the literature that the main effect of the PSF is to bring light from the center to the outskirts of the galaxies and, therefore, change their properties \citep[e.g.][]{Capaccioli1983, Trujillo2016}. That produces a redder ICL, a change in the shape of the spectral energy distribution (SED) and, therefore, a change in the derived ages and metallicities. To compensate for this effect, we PSF-matched all the images to the worst resolution one: F160W. As we need to characterize the PSF accurately up to a large distance \citep[][]{Sandin2014}, we built the PSFs of each band in two steps. The inner region is created using the software \texttt{Tiny Tim}. Based in the shapes of the inner profiles, the outer parts are assumed to follow an exponential behaviour. Once the PSFs of each band are built, we created a kernel for each image such as the convolution of that kernel with our images results in a new image matching the resolution of F160W. Details on the full process can be found in Section \ref{psf}.

\subsection{Background determination and subtraction} \label{back}

An accurate background determination of the HFF data is very difficult, particularly in the IR bands, as the ICL almost fills the entire image. Furthermore, a preliminary exploration of the images showed that most of them had their background over subtracted during the data reduction process. To correct for such artificial sky, we add a constant background to the images.
To do that, we measured the sky in each band of each cluster in $\sim 30$ apertures of $r$ = $25$ pix ($1\farcs5$), located farther away from the cluster centre (i.e. $>200$ kpc for the clusters with two brightest cluster galaxies (BCGs) and $>350$ kpc for clusters with one BCG; see below for the BCG definition). In addition, we avoid any source or diffuse light so we can estimate a sky value for our images without contamination. To test the gaussianity of our sky distributions, we performed Kolmogorov-Smirnov tests that indicate that a Gaussian distribution is compatible with the data, with p-values $>0.05$.

To measure the ICL accurately, it is crucial to masks all sources that might contaminate the diffuse light. The majority of the HFF clusters are merging or have undergone a recent merger \citep[][]{Lotz2017}, hence in some cases the choice of which galaxy is the BCG is unclear. Consequently, when the difference in the magnitudes of the two most massive galaxies is small, we chose both as BCGs. The clusters where we have two BCGs are: A2744, M0416 and A370. Once identified our BCGs, we proceeded to mask all the galaxies of the cluster as well as foreground and background sources. The masking process is detailed in Section \ref{masks}. The masked images are presented in Fig. \ref{app:masks}.

The purpose is to study the properties of the stellar populations of the ICL down to the faintest surface brightness possible. Therefore, to estimate the surface brightness limits down to where we can explore the ICL, we calculated the r.m.s of the images on boxes of $3\times3$ arcsec$^2$ located on the sky for each of the clusters. The surface brightness limits we provide correspond to $3\sigma$ detections above the sky in the PSF-matched images. The surface brightness limits are listed, in each band and for each of the clusters in Table \ref{table:table1}.


\begin{table*}
 \centering
 \tabcolsep 2.5pt
  \begin{tabular}{@{}lccc|ccccccc@{}}
	Cluster 		 & Redshift &          R.A. (2000)      &      Dec (2000)        &    \multicolumn{7}{c}{SB limits (mag/arcsec$^2$)}  \\ 
	            		 &    (z)       &  (hh:mm:ss) &  (dd:mm:ss) &      F435W          &     F606W            &  F814W             &     F105W          &      F125W           &   F140W         &   F160W \\ \hline 
 	Abell  2744 		  & $0.308$ &  00:14:21.2  &  -30:23:50.1  &  $31.53\pm0.09$ & $31.33\pm0.12$ & $31.71\pm0.10$ & $31.91\pm0.12$ & $31.64\pm0.09$ & $31.65\pm0.11$ & $30.24\pm0.12$ \\
	MACSJ0416.1-2403   & $0.396$ &  04:16:08.9  &  -24:04:28.7  &  $31.71\pm0.11$ & $31.88\pm0.14$ & $31.93\pm0.14$ & $31.97\pm0.11$ & $31.85\pm0.12$ & $31.78\pm0.11$ & $30.32\pm0.12$ \\
	MACSJ0717.5+3745  & $0.545$ &  07:17:34.0  &  +37:44:49.0 &  $31.27\pm0.13$ & $31.40\pm0.09$ & $31.61\pm0.10$ & $31.85\pm0.11$ & $31.45\pm0.16$ & $31.58\pm0.11$ & $30.16\pm0.13$ \\
	MACSJ1149.5+2223  & $0.543$ &  11:49:36.3  &  +22:23:58.1 &  $31.10\pm0.11$ & $31.56\pm0.09$ & $31.56\pm0.11$ & $31.70\pm0.12$ & $31.31\pm0.12$ & $31.30\pm0.14$ & $30.19\pm0.15$ \\
        Abell S1063                & $0.348$ &  22:48:44.4  &  -44:31:48.5  &  $31.37\pm0.15$ & $31.36\pm0.10$ & $31.68\pm0.10$ & $31.72\pm0.12$ & $31.51\pm0.11$ & $31.65\pm0.11$ & $30.25\pm0.15$ \\
        Abell 370                     & $0.375$ &  02:39:52.9  & -01:34:36.5   &  $31.17\pm0.16$ & $31.42\pm0.15$ & $31.71\pm0.11$ & $31.84\pm0.10$ & $31.47\pm0.11$ & $31.53\pm0.11$ & $30.17\pm0.13$ \\\hline
  \end{tabular} 
 \caption{Summary of the main properties of the six HFF clusters. The surface brightness limits (3$\sigma$ above the sky) are calculated in boxes of $3\times3$ arcsec$^2$. These limits have been obtained for the images at the same spatial resolution as the F160W image.}\label{table:table1}
 \vspace{1cm}
\end{table*}


\section{Results}\label{results}

\subsection{Spectral Energy Distributions}

\begin{figure}
 \begin{center}
  \includegraphics[scale=0.5]{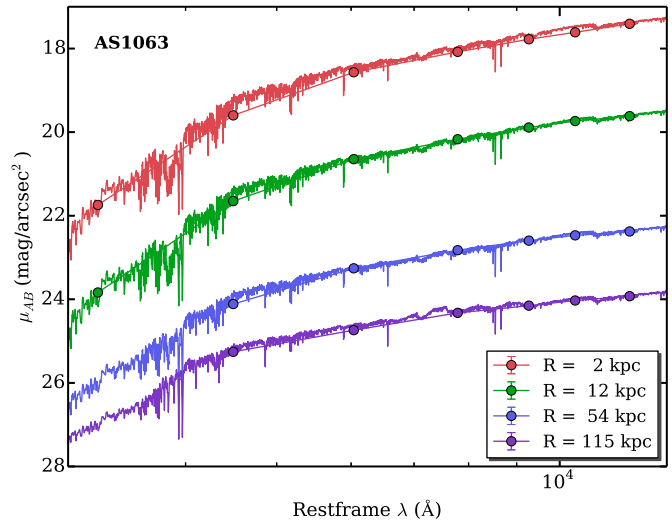}  
  \caption{Example of SEDs at four radial distance for one of the HFF clusters, AS1063. The filled circles represent the SED derived from the images and redshifted to their restframe wavelength, while the solid lines represent the best fitting \citet[][]{Vazdekis2016} models at each radius. The errors of the photometry are smaller than the size of the circles. The surface brightness, vertical axis, is corrected by cosmological dimming. The bluest band, F435W, in the SEDs at $R $= $54$ and $R$ = $115$ kpc shows a level of background contamination above $50\%$, therefore we are not using it for the fits.} 
 \label{fig:seds}
 \end{center}
\end{figure}

The goal of this work is to study the stellar populations of the HFF clusters from the center of their BCG(s) to the outer parts of the clusters. To that end, we derived the radial SEDs of the clusters in $16$ logarithmic spaced bins from $0$ to $200$ kpc from the BCG(s). The distance of each pixel on the images is computed as the elliptical distance to its nearest BCG, where the morphological parameters of these galaxies are given by SExtractor (see \citealt{MT14} for further details). For each radial bin, the surface brightness was obtained averaging the pixel values. The errors of the SED values are a combination from the photometric errors and the zeropoint uncertainties. The photometric errors are drawn from jackknife resampling, i.e. repeating the photometry in a subsample of the data for each bin. The number of subsamples taken was $500$. 
We corrected our data for Galactic extinction \citep[][]{Schlafly2011} using the \citet[][]{Cardelli1989} extinction law. We also corrected for cosmological dimming. 
As each image has a different depth, we decided to use a simple method to determine up to which radius our photometry is reliable. For that, we determine the contamination of sky background pixels in each spatial bin. To do that, we compared the observed distribution of counts in each spatial bin with the sky background distribution (see Section \ref{back} for the determination of the background). The photometry in each band is estimated until the level of background contamination is more than $50\%$. As we expect old or intermediate ages for the ICL (Paper I), it should be fainter in the bluer filters. This implies that our completeness is ultimately limited by the noise in the observed optical filters. Eliminating the optical filters from our analysis increases the degeneracy between age and metallicity, i.e. the errors on the determination of those quantities. Fig. \ref{fig:seds} shows an example of different SEDs at different radial distances. We are not plotting the bluest band, F435W, for the SEDs at $R $= $54$ and $R$ = $115$ kpc as their level of sky background contamination is more than $50\%$.

\subsection{Methodology} \label{method}

Taking advantage on the large wavelength range of our data set, we fitted single stellar population (SSP) models to explore whether it is possible to detect stellar population gradients from the center of the BCG(s) to the outer parts of the cluster. Note that considering an SSP at each radius is a rough assumption since we expect the ICL to be formed by the accretion of a variety of galaxies, especially at large distances from the center of the cluster (see Paper I). Nonetheless, and for simplicity, we follow the approach as described in Section 3.2.1 in \citet{Montes2014a}, and briefly described here. 

In this work, we use the UV-extended E-MILES SSP models from \citet{Vazdekis2016} for the Padova 2000 isochrones. The E-MILES models cover the spectral range $1680-50000$ \AA{}, and consist of $7$ metallicities in the range $-1.79\leq\,$[Fe/H]$\,\leq +0.26$ and $50$ ages from $0.03$ to $17.8$ Gyr, for a suite of initial mass function (IMF) types with varying slopes. To diminish the uncertainties due to width of the steps in age and metallicity, we expanded the grid of models with $200$ metallicities and $200$ ages linearly interpolating the original SSPs. Our choice of IMF is a \citet{Salpeter1955} IMF, i.e. a unimodal IMF with a slope of $1.3$. For each cluster, the maximum age allowed in our fits is the age of the universe at the given redshift.

The observed SEDs are compared with the model SSPs to obtain information of the stellar populations at each radius (see Fig. \ref{fig:seds}). For this, we first redshifted the model to the redshift of the cluster. Then, we convolved the redshifted model with the filter response of our photometric filters to retrieve synthetic photometry for comparison. We used a $\chi^2$-minimization approach to obtain the best fit model to our data \citep[Eq. 2 in][]{Montes2014a}, as well as the 1-$\sigma$ confidence levels. As the parameters to fit are three: age, metallicity and luminosity, the number of spectral bands required for fitting are at least four.

We run the fitting procedure for each of the jackknife realizations of the SEDs. The final ages and metallicities are the median of the ages and metallicities of the realizations. The errors are the median errors of each of the fits divided by the square root of the number of realizations.

\subsection{Age and metallicity gradients}\label{agemets}

In Fig. \ref{fig:fig3}, we present the age and metallicity profiles for each of the HFF clusters. In the left panels, we overplot the colour coded distance bins to the masked image of the cluster. Middle and right panels are, respectively, the age and metallicity profiles up to $200$ kpc, depending on the cluster. 

As we mentioned in Paper I, the definition of ICL is controversial as it is unclear how to disentangle, if possible, between the BCG(s), especially the outskirts, and the ICL. Several studies attempt this by defining a surface brightness threshold \citep[e.g.][]{Feldmeier2004, Mihos2005, Krick2006, Krick2007, Burke2015}. In our case, we define ICL as the light beyond $> 50 $ kpc \citep[see][]{Gonzalez2005, Toledo2011}. Although this choice of radius is not perfect, as it could include a fraction of light from the outskirts of the BCG(s), it allows us to compare with previous spectroscopic studies that derive ages and metallicities for the ICL at those distances \citep[e.g.][]{Coccato2010, Melnick2012, Edwards2016}. Furthermore, \citet[][]{Presotto2014} fit a de Vaucouleurs profile to the BCG of MACSJ1206.2-0847 ($z\sim0.44$). They found that at $R>40$ kpc there is an excess of light with respect to the r$^{\frac{1}{4}}$ fit. They identify this deviation from a de Vaucouleurs profile as the signature of the ICL. 
The region we consider ICL is highlighted as an orange shaded area in Fig. \ref{fig:fig3}. According to the middle panels in Fig. \ref{fig:fig3}, there is a continuos negative age gradient from the centre of the cluster to their outskirts. That implies that the outskirts of the clusters are younger than the centre of the BCG(s). That is true also for the metallicity gradients (right panels in Fig. \ref{fig:fig3}) where we can see that the stellar populations become more metal-poor as the distance from the BCG(s) increases. 
The distances, ages and metallicities values are listed in Table \ref{table:agemet}.

\begin{figure*}
\centering
\begin{subfigure}[t]{\textwidth}
\includegraphics[width=\textwidth]{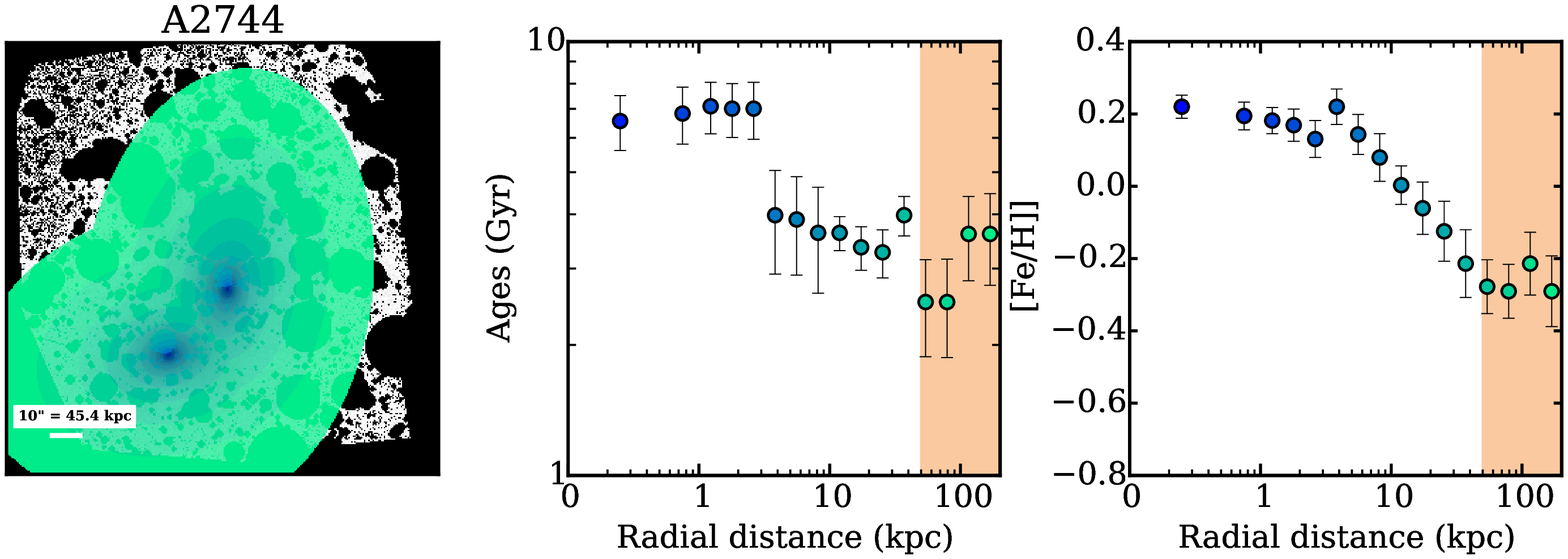} 
\end{subfigure}
\begin{subfigure}[t]{\textwidth}
\includegraphics[width=\textwidth]{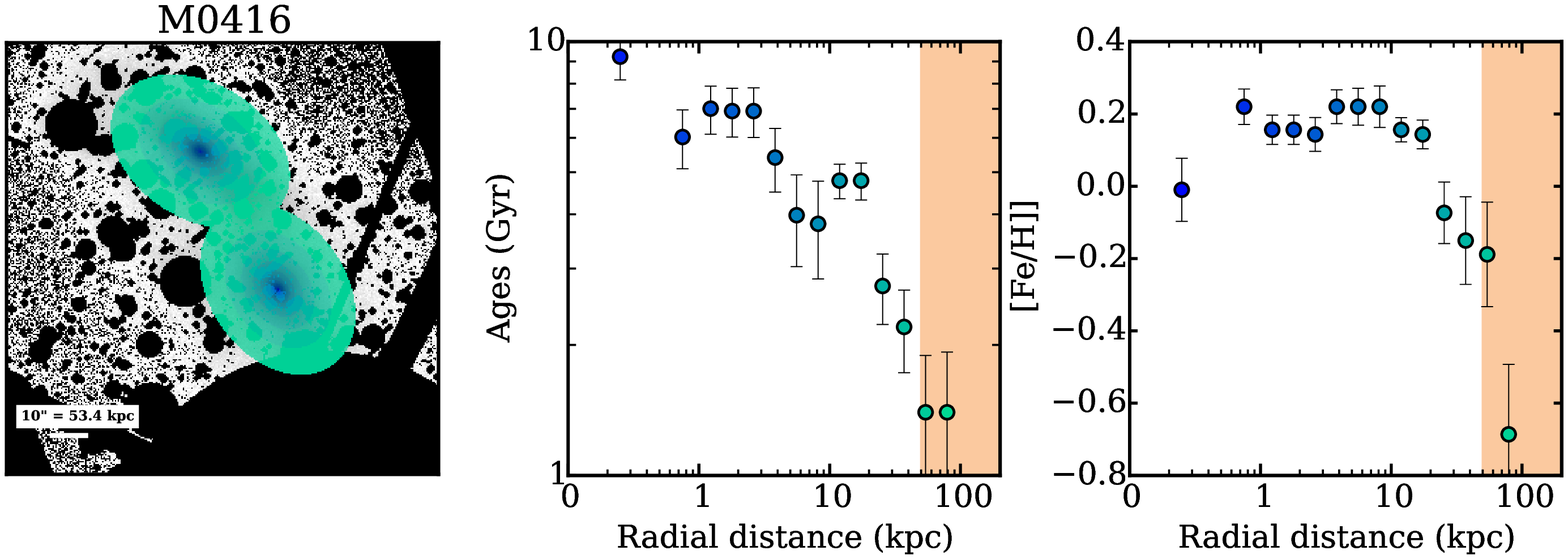} 
\end{subfigure}
\begin{subfigure}[t]{\textwidth}
\includegraphics[width=\textwidth]{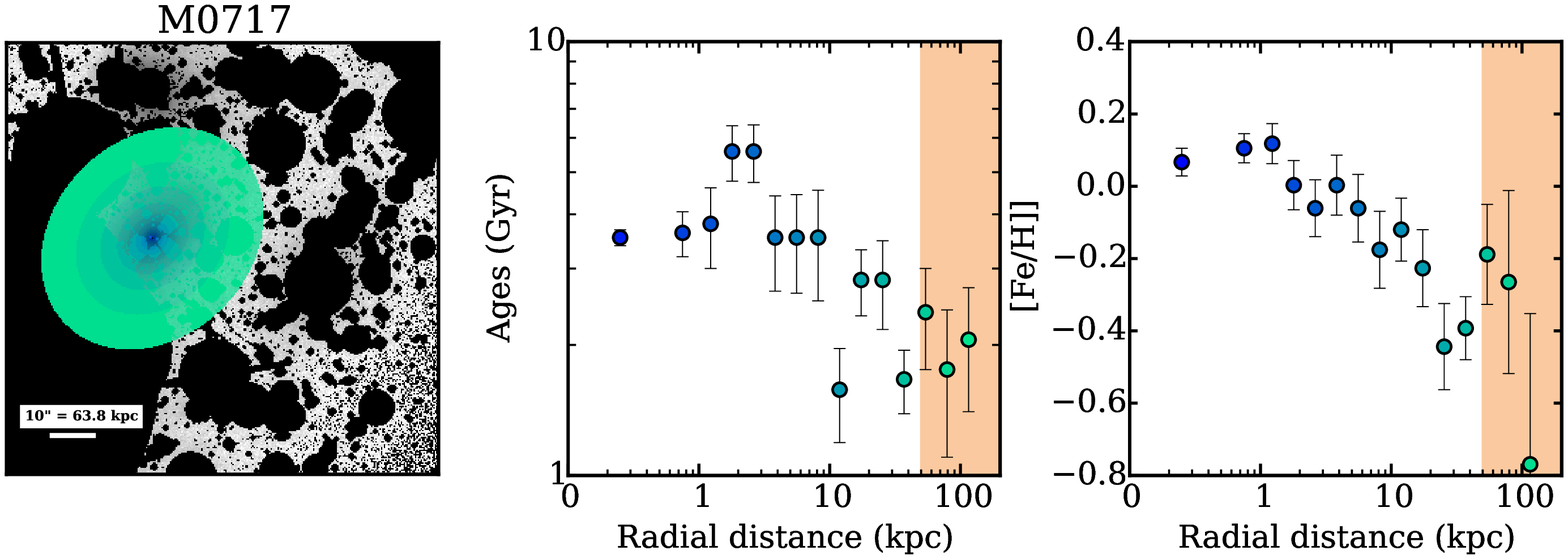} 
\end{subfigure}
 \caption{Gradients of age and metallicity as a function of radial distance to the BGC(s) of the clusters. Left panels show the image of the cluster in the F160W filter and overplotted are the different spatial regions in which the SEDs are measured. The central and right panels are the age and metallicity radial gradients derived from the fitting to the SEDs as described in Section \ref{method}. We also marked in orange the region of the ICL ($R>50$ kpc). The colours of the spatial regions in the left panels correspond to the colours of the circles in the age and metallicity gradients. } 
 \label{fig:fig3}
\end{figure*}

\begin{figure*}\ContinuedFloat
\begin{subfigure}[t]{\textwidth}
\includegraphics[width=\textwidth]{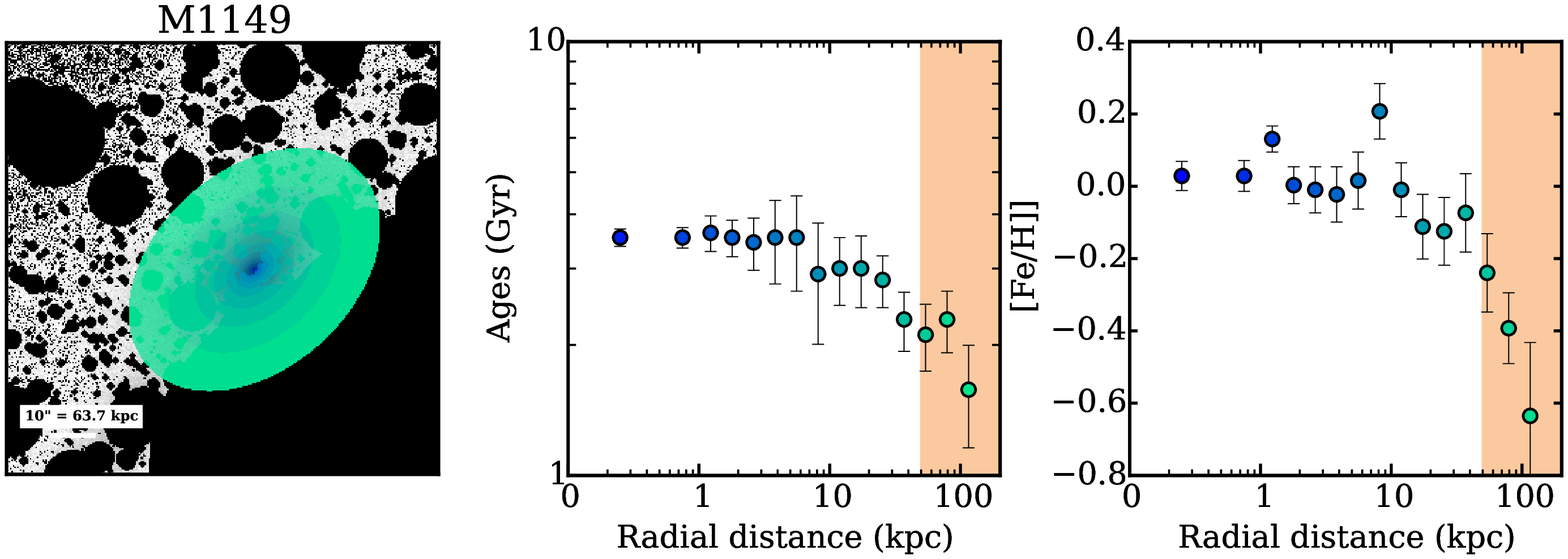} 
\end{subfigure}
\begin{subfigure}[t]{\textwidth}
\includegraphics[width=\textwidth]{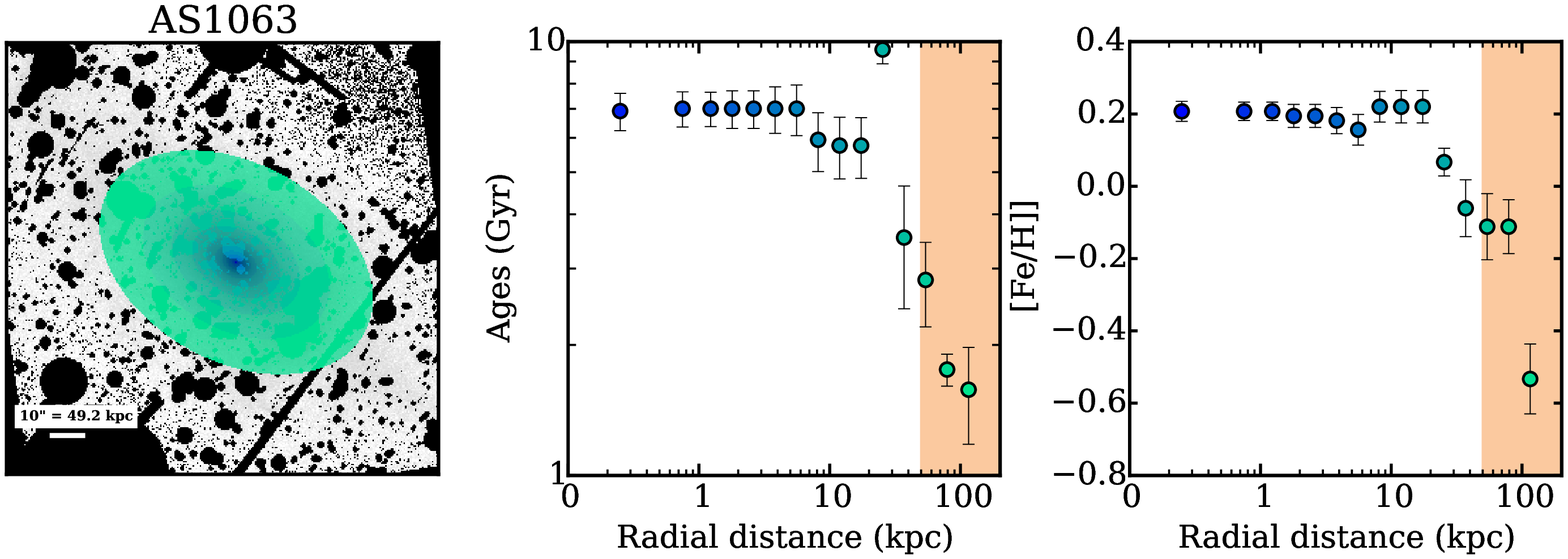} 
\end{subfigure}
\begin{subfigure}[t]{\textwidth}
\includegraphics[width=\textwidth]{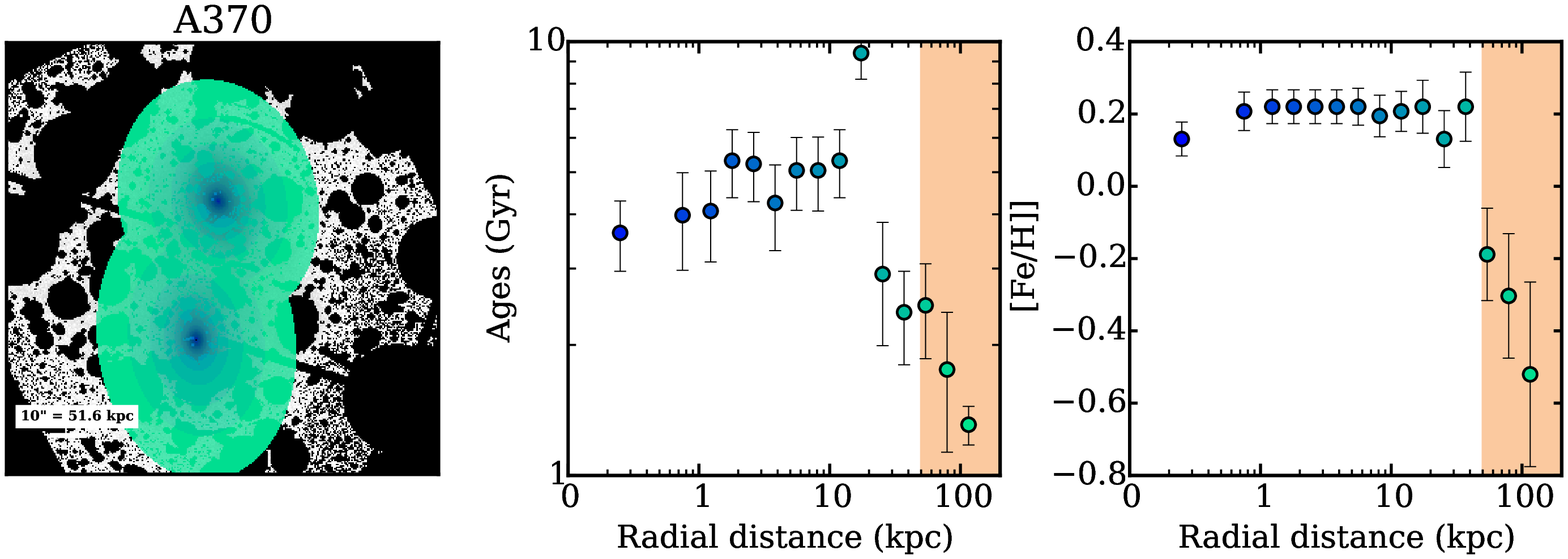} 
\end{subfigure}
  
 \caption{Continued} 
\end{figure*}

\subsubsection{The HFF clusters}\label{clusters}

\paragraph*{Abell 2744} 
Abell 2744 is the nearest cluster of the HFF survey (z = $0.308$) also known as "Pandora's Cluster". It comprises $4$ different mass substructures \citep{Merten2011} and a complex velocity structure suggestive of a merging system \citep[][]{Braglia2007, Merten2011}. The HFF survey imaged the most massive southern structure, the core \citep{Merten2011}. Its velocity dispersion is $\sigma$ = $1497\pm47$ km/s$^{-1}$ \citep{Owers2011}.
In Fig. \ref{fig:fig3}, it is shown that this cluster presents a negative gradient in age, ranging from $\sim7$ Gyr in the centre of the BCGs to $\sim3.5$ Gyr in the ICL, the orange shaded region. Despite the differences in methodology and stellar population models, this agrees with our previous results for this cluster (see Paper I), where we found that the ICL was between $6\pm3$ Gyr younger than the center of the most massive galaxies of the cluster.
For the metallicity, we find a value of [Fe/H]$\sim -0.3$ for the ICL, Z $\sim 0.008$, while for the BCGs is [Fe/H] $\sim 0.22$ (Z$\sim 0.03$). This metallicity value is slightly lower than what we found in Paper I but in agreement within errors.

\paragraph*{MACSJ0416.1-2403} 
This object is an elongated merging cluster \citep[][]{Mann2012}, also part from the CLASH survey \citep[][]{Postman2012}. It has a velocity dispersion of $779\pm22$ km/s$^{-1}$ \citep[][]{Ebeling2014}. One of the most prominent features in the image of the cluster is the presence of a bright star within $1$ arcminute from the cluster core. Although the FF images avoid the center of the star, the halo and spikes can be clearly seen. Therefore, we decided to aggressively mask that region (see Fig. \ref{app:masks}). 
The ages range from $\sim 8$ Gyr in the core of the BCGs to $1.4\pm0.5$ Gyr in the region of the ICL ($>50$ kpc) giving a difference of $\sim 6$ Gyr. The metallicities range from [Fe/H]$\sim 0.22$ to $\sim-0.5$. 

\paragraph*{MACSJ0717.5+3745}
This cluster is the farthest (z = $0.545$, \citealt{Edge2003}), one of the most massive and the strongest lenser of all the clusters in the HFF sample \citep[][]{Lotz2017}. The velocity dispersion is $1660\pm125$ km/s$^{-1}$ \citep[][]{Ebeling2007}. It has some stars and a foreground galaxy in the field of view (FOV). The extended outskirts of this foreground galaxy is a source of contamination of the ICL, therefore we decided to mask most of the left side of the image (see Fig. \ref{app:masks}). The ages range from $\sim 4$ Gyr in the centre of the cluster to $\sim 2$ Gyr in the outer parts, a difference of $2$ Gyr. The metallicities range from [Fe/H]$\sim 0.12$ to [Fe/H] $\sim - 0.5$.

\paragraph*{MACSJ1149.5+2223}
A cluster at z=$0.543$, MACSJ1149.5+2223 is an X-ray elongated cluster with a complex merger history \citep{Kartaltepe2008, Zitrin2009, Lotz2017}. Its velocity dispersion is $1840\pm150$ km/s$^{-1}$ \citep[][]{Ebeling2007}. This cluster also presents two bright stars and a foreground galaxy near the core of the cluster, therefore the lower right part of the image is masked to prevent contamination (see Fig. \ref{app:masks}). The ages range from $\sim4$ Gyr to $\sim 2$ Gyr in the ICL region, a difference of $2$ Gyr. The metallicities go from [Fe/H] = $0.22$ to $\sim-0.4$ for the ICL.

\paragraph*{AS1063}
This is massive cluster with significant substructure and a velocity dispersion of $1840\pm190$ km/s$^{-1}$ \citep[][]{Gomez2012, Gruen2013} at z=$0.348$ \citep[][]{Gomez2012}. It is the most relaxed of the selected HFF clusters \citep[][]{Lotz2017}. The ages range from $\sim7$ Gyr in the centre of the cluster to $\sim 1.7$ Gyr in the ICL region, a difference of $\sim 5$ Gyr. The metallicities go from [Fe/H] = $0.21$ to $\sim-0.3$ for the ICL.

\paragraph*{Abell 370}
Abell 370 is a very well studied lensing cluster \citep[see references in ][]{Lotz2017} at z = $0.375$ \citep[e.g.][]{Struble1999}. Its total velocity dispersion is $1170\pm100$ km/s$^{-1}$ \citep[][]{Dressler1999}. The ages range from $\sim5$ Gyr in the centre of the BCGs to $\sim 1.5$ Gyr in the ICL region. The metallicities go from [Fe/H] = $0.22$ to $\sim-0.4$ for the ICL.

\subsection{The contribution of the ICL to the total amount of light} \label{fractions}

Recently, both observations \citep[][]{Burke2015} and simulations \citep[][]{Rudick2011, Contini2014} have suggested that there is a strong evolution in the fraction of light contained in the ICL at later times, $z<1$. \citet[][]{Burke2015} showed that the most dramatic evolution in the fraction of this component starts at $z\sim0.5$ \citep[see also][]{Krick2007}. Therefore, the redshift range spanned by the HFF clusters, $0.3<z<0.55$, appears as an interesting epoch to explore whether it is the onset of the ICL in galaxy clusters.
In order to investigate this, we derived the fraction of light contained in the ICL in the following fashion. We measured the ICL flux from the clusters images applying an ICL threshold of  $\mu_{V}$ = $26$ mag/arcsec$^2$ to be able to compare with previous studies and simulations. To measure the total cluster light, we derived again the masks for each cluster but this time not including all the galaxies belonging to the cluster. The mask was constructed using the available spectroscopic and photometric redshifts provided by the HFF team\footnote{http://www.stsci.edu/hst/campaigns/frontier-fields/FF-Data} \citep[][]{Lotz2017}  when available. We also used grism redshifts for AS1063 and A370 provided by the GLASS team\footnote{https://archive.stsci.edu/prepds/glass/} and NED\footnote{http://ned.ipac.caltech.edu/} photometric redshifts for A2744 and A370. We identify a galaxy as a member of the cluster if its redshift does not depart from the redshift of the cluster by more than $\Delta z $= $0.05$, to account for the photometric redshift errors. As the brighter cluster members have spectroscopic or grism redshift available, the choice of $\Delta z$ only affects low mass galaxies. We tried with different values of $\Delta z$ and the changes were insignificant (if any). It is worth stressing that our estimation of the total stellar light of the cluster depends on its redshift. In other words, the higher the redshift the larger the number of low mass galaxies we are missing. We discuss how this incompleteness can affect our results in Appendix \ref{completeness}.

To construct the mask, we assigned a redshift value to the pixels of the images using the segmentation map given by the SExtractor runs (see Section \ref{masks}). Then, we masked the area subtended by those objects whose redshifts do not correspond to the cluster and also those without redshift. Finally, the image in the rest-frame V-band for each cluster, necessary for applying the ICL criteria, is computed by interpolating among the observed HST images.

\begin{figure}
 \begin{center}
  \includegraphics[scale=0.45]{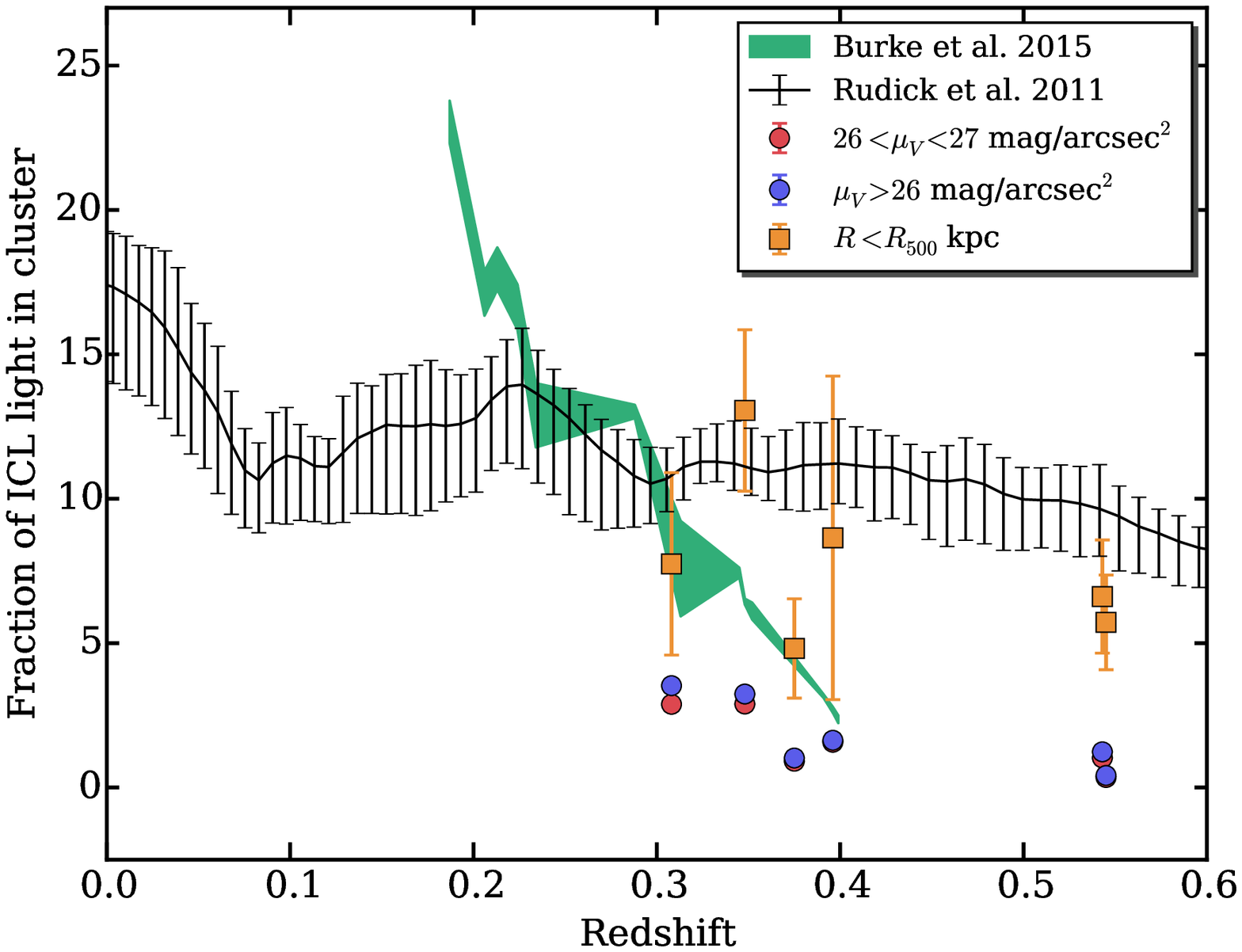}  
  \caption{Fraction of light in the V-band in the ICL component as a function of redshift. Red filled circles correspond to a slice in surface brightness in the V-band from 26 to 27 mag/arcsec$^2$. Blue filled circles correspond to all the light fainter than $\mu_{V}$= $26$ mag/arcsec$^2$ while the orange squares correspond to the fraction of ICL light for $R<R_{500}$ kpc. The green polygon are the fractions derived in \citet[][]{Burke2015}, for $13$ clusters from CLASH between $\mu_B$ = $25$ and $26$ mag/arcsec$^2$. The black line is the prediction of \citet[][]{Rudick2011} for the fraction of ICL measured with $\mu_V > 26$ mag/arcsec$^2$. } 
 \label{fig:fig4}
 \end{center}
\end{figure}

We present the fraction of total cluster light contained in the ICL of the HFF clusters in the V-band as the blue and red filled circles in Fig. \ref{fig:fig4}. Red filled circles in Fig. \ref{fig:fig4} indicate the fraction of the ICL contained in a slice in $\mu_{V}$ from 26 to 27 mag/arcsec$^2$ ($\sim3 \sigma$ above the background). To account for the possible bias introduced by imposing a faint-end limit on the ICL \citep[see also][]{Rudick2011, Burke2015} we also measure the fraction of ICL by including all the pixels fainter than our ICL threshold $26$ mag/arcsec$^2$ (blue filled circles in Fig. \ref{fig:fig4}). When doing that, we find an average increase in the ICL fraction is $\sim13\%$. In Fig. \ref{fig:fig4}, we compared with the observational data in \citet[][]{Burke2015}. They derived ICL fractions for $13$ CLASH clusters below a surface brightness threshold in the B-band of 25 mag/arcsec$^2$ and above 26 mag/arcsec$^2$. The threshold in the B-band is equivalent to a threshold of $24.3$ mag/arcsec$^2$ in the V-band, assuming a colour of B-V = $0.7$ \citep[][]{Vazdekis2016}, for an age of $2$ Gyr and [Fe/H] = $-0.4$, similar to our derived ages and metallicities for the ICL. As their threshold is brighter, the ICL fractions in \citet[][]{Burke2015} include more light from the inner parts of the cluster than in our case, i.e. we reach further out from the inner parts of the cluster. The measured fractions for the HFF clusters for both $26<\mu_V<27$ mag/arcsec$^2$ and $\mu_V>26$ mag/arcsec$^2$ are listed in Table \ref{table:table2}.

The slope of the \citet[][]{Burke2015} points shows a steep increase in the fraction of ICL with redshift indicating that the build-up of this component is fairly rapid with decreasing redshift. In our data, we see a slight increase on the ICL fraction with decreasing redshift, as predicted by simulations \citep[][]{Rudick2011, Contini2014} although not as steep as in \citet[][]{Burke2015}. However, we are exploring a narrow redshift range with respect to \citet[][]{Burke2015}, therefore any conclusions must be taken with caution.

We also plotted in Fig. \ref{fig:fig4} the redshift evolution of the ICL fraction from \citet[][]{Rudick2011} simulations for a cluster of mass $M$=$0.84\times10^{14} M_{\odot}$, lower than the mass of the HFF clusters. Their data points are measured using the same surface brightness threshold than in this work: i.e. $\mu_V>26$ mag/arcsec$^2$. However, they include all the light brighter than $\mu_V$ = $33$ mag/arcsec$^2$ \citep[][]{Rudick2006}, $6$ magnitudes fainter than our fainter lower limit ($27$ mag/arcsec$^2$). Consequently, the difference between the fraction of ICL from their simulations and our observations could be mainly caused by the light we are missing in our observations. We found an average increase of $13\%$ in the ICL fraction when we lower our faint limit to include all the light below $26$ mag/arcsec$^2$. In the same way, \citet[][]{Burke2015} observed an increase of $40\%$ when they lowered their faint limit from $25.5$ to $26$ mag/arcsec$^2$ in the B-band. Unfortunately, there is no estimate of the amount of light we could be missing.

The transition between BCG and ICL happens smoothly, making it difficult to disentangle between both components. To overcome this issue we have used a surface brightness threshold approach, which is unable to account for the amount of ICL that (in projection) is above the BCG. Therefore, to account for it, we decided to perform a linear fit to the surface brightness of each cluster in the restframe V-band for the ICL region, i.e. $R>50$ kpc, as a way to describe this component. The functional form for the fit is suggested by the relatively flatness of the ICL profiles \citep[see][]{Krick2007, Cooper2015}. Using the fit as the ICL profile, we have evaluated the fraction of ICL for three different apertures: $50$ kpc$<R<R_{limit}$, $R<R_{limit}$ and $R<R_{500}$ (the latter are the orange filled squares in Fig. \ref{fig:fig4}). $R_{limit}$ is defined as the radius in which the number of bands for the SSP fits are at least four. They can be found in Table \ref{table:agemet} as the last bin with age and metallicity for each cluster. $R_{500}$ is the radius where the mean mass density exceeds the critical density by a factor of $500$. They are taken from \citet[][]{Mantz2010, Maughan2012, Ehlert2013} and \citet[][]{Sayers2013}, and are on average $\sim1.5$ Mpc. There are two clusters, M0717 and M1149, whose fractions change radically when we change the outer radius. 
Those two clusters are significantly less concentrated than the rest of the FF clusters. Therefore, as we go farther from the centre of the cluster, the quantity of light in the ICL diminishes, but we are including more galaxies, more stellar mass, of the cluster resulting in an overall decrease of the fraction of light in the ICL.


The ICL fractions derived for the different apertures are also listed in Table \ref{table:table2}. The average fraction of ICL is $\sim7\%$ from the centre of the BCG(s) to the outer parts of the cluster, $R<R_{500}$. None of these definitions of ICL support a redshift evolution of the fraction of ICL, although the redshift range we are exploring is not large enough to draw any strong conclusion.

\begin{table*}
 \centering
 \tabcolsep 2.5pt
  \begin{tabular}{@{}lccccc@{}}
	Cluster 		        &\multicolumn{5}{c}{\% of light in ICL}  \\ 
	                                  &  $26<\mu_{V}<27$  & $\mu_v > 26 $   &   $50< R< R_{limit}$ & $R<R_{limit}$   &    $R<R_{500}$ \\ \hline
 	Abell  2744        	 & $2.88\pm0.07$       & $3.53\pm0.13$   &   $7.1\pm3.2$   &    $9.1\pm3.7$    &    $7.7\pm3.1$    \\
	MACSJ0416.1-2403  & $1.57\pm0.03$	  & $1.63\pm0.04$   &   $5.6\pm2.3$   &    $11.5\pm3.7$   &    $8.6\pm5.6$   \\
	MACSJ0717.5+3745 & $0.35\pm0.01$       & $0.41\pm0.02$   &   $18.9\pm5.3$  &    $27.4\pm7.6$   &    $5.7\pm1.6$   \\
	MACSJ1149.5+2223  & $1.03\pm0.03$       & $1.23\pm0.05$   &   $19.5\pm5.7$  &    $29.6\pm7.8$   &    $6.6\pm2.0$    \\
        Abell S1063                & $2.89\pm0.04$       & $3.24\pm0.06$    &   $15.4\pm3.3$  &    $23.5\pm4.6$  &    $13.1\pm2.8$    \\
        Abell 370                     & $0.91\pm0.02$       & $1.02\pm0.03$   &   $6.9\pm2.5$    &    $10.7\pm3.5$   &    $4.8\pm1.7$    \\\hline

  \end{tabular} 
 \caption{ Fractions of ICL light measured in the V-band for the HFF clusters. The  first two columns are the fractions assuming a surface brightness threshold to define the ICL region, while in the other columns the ICL component is thought to follow a linear profile, derived from a fit to the surface brightness profile in the V-band (R>$50$ kpc). $\mu_V$ is in mag/arcsec$^2$. }\label{table:table2}
 \vspace{1cm}
\end{table*}


\subsection{Slopes of the stellar mass density profiles of the ICL}\label{slopes}

The possibility that the stellar halo or ICL properties might provide information about the assembly history of the parent dark matter halo has motivated substantial effort to study the outskirts of galaxies and clusters. These diffuse components are thought to be direct evidence of hierarchical growth in the cold dark matter scenario in the sense that more massive dark matter haloes will accrete more, and more luminous, satellites \citep[e.g.][]{Gao2004}. Recently, \citet[][]{Pillepich2017b} analysed a sample of $4000$ galaxies in the IllustrisTNG simulation \citep[][]{Vogelsberger2014, Pillepich2017a, Weinberger2017}. They found that there is a strong correlation between the slope of the density profile of the stellar halo and the total mass of the system spanning a wide range of halo masses (M$_{200}$ = $\sim10^{13}$ to $10^{15}$ M$_\odot$).

\begin{figure}
 \begin{center}
  \includegraphics[scale=0.45]{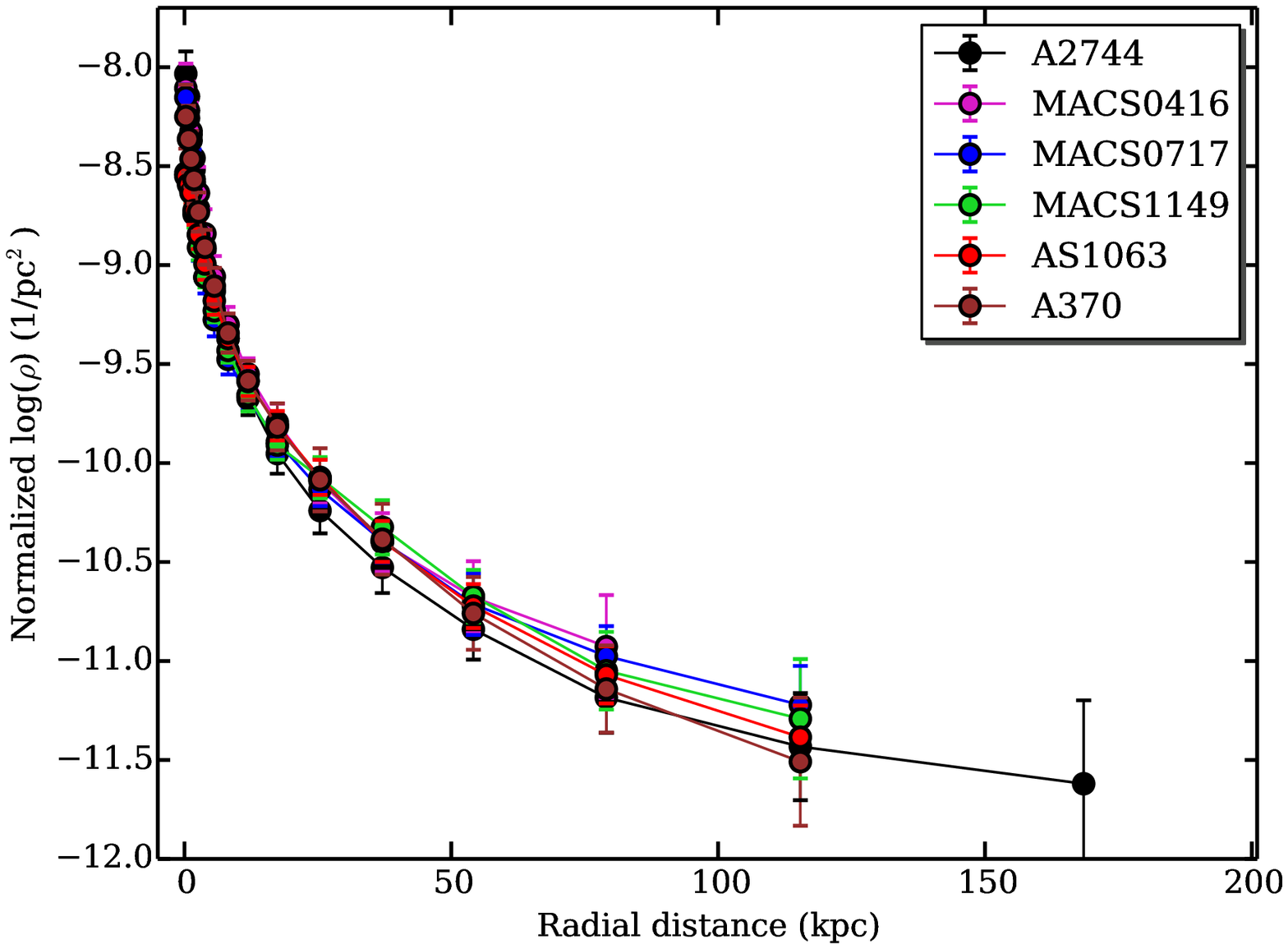}  
  \caption{Mass normalized density profiles for the HFF clusters. } 
 \label{fig:fig5}
 \end{center}
\end{figure}

To explore this, we have derived the stellar mass density profiles for the HFF clusters. To this end, we have followed the procedure we used in Paper I \citep[see also][]{Bakos2008}.  Applying Equation 1 and 2 from Paper I,  we linked the observed surface brightness in the restframe z-band to a mass to light (M/L) ratio. The M/L ratio was derived from the prescriptions given by \citet[][]{Bell2003}, using an i-z colour for a \citealt[][]{Salpeter1955} IMF. Then, the radial profiles were derived using the same distance bins as in the case of the SEDs. Fig. \ref{fig:fig5} shows the density profiles of the HFF cluster normalized by their total stellar mass ($R<R_{limit}$) for comparison. The differences between the normalized profiles of the clusters are small. As galaxies merge into the BCG, one will expect to find a less steep profile as cosmic time progresses. However, we do not see any redshift dependence of the slope of the density profile with redshift, pointing that redshift might not be the driver of this change (at least for these massive clusters). Additionally, relaxed clusters are dynamically older clusters that have already been through significant mergers. Consequently, one could expect to see a shallower density profile in relaxed clusters. However, we do not see any evidence of a difference in the most relaxed cluster in the sample, AS1063.

The slopes of the stellar mass density profiles are computed fitting a linear relationship in logarithmic space to the ICL component ($R>50$ kpc, see Fig \ref{fig:dens}). \citet[][]{Pillepich2017b} measured the 3D spherically averaged density profiles for the stars. To translate between our observed 2D slope to their 3D slope we used Equation 5 in \citet[][]{Stark1977}. The total masses of our clusters ($M_{200}$) were computed using the relationship between $M_{200}$ and velocity dispersion given by \citet[][]{Munari2013}. The velocity dispersions for each of the clusters are given in Sec. \ref{clusters}.

\begin{figure}
 \begin{center}
  \includegraphics[scale=0.45]{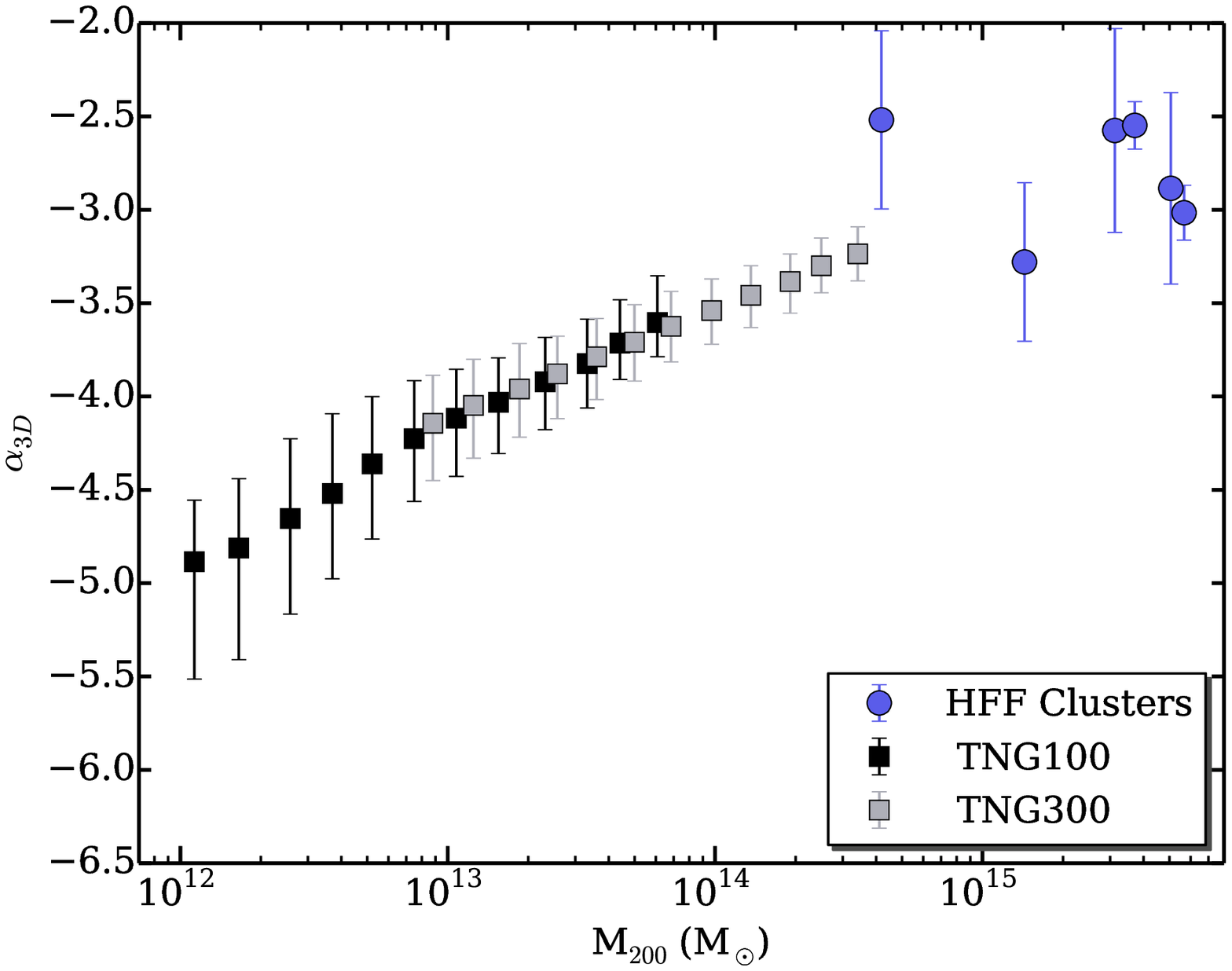}  
  \caption{Slope of the stellar mass density profile for $R>50$ kpc vs. the mass of the halo (M$_{200}$). Blue filled circles are our measured slopes for the HFF clusters. Black and grey squares correspond to the IllustrisTNG simulations taken from \citet[][]{Pillepich2017b}, for a volume of $\sim100^3$ Mpc$^3$ and $\sim300^3$ Mpc$^3$. They measured the stellar mass density profiles between $30$ kpc to $2R_{hm}$.} 
 \label{fig:fig6}
 \end{center}
\end{figure}

We warn the reader that the comparison with \citet[][]{Pillepich2017b} has to be taken with caution as some approximations were made that might not be accurate enough. For instance, we are comparing slopes derived at different radii. To measure the slope, we took the profile at $R>50$ kpc ending in most cases around $\sim120$ kpc, while \citet[][]{Pillepich2017b} fits the slope in a wider range ($30$ kpc $<R<2R_{hm}$, with $R_{hm}$ being the 3D half mass radius). In other words, we assumed that the slope of the stellar mass density profile is not changing at larger radius \citep[a reasonable assumption given the profiles in Fig. 1 in][]{Pillepich2014}. The HFF clusters are more massive than the objects explored in \citet[][]{Pillepich2017b}. Nevertheless, we can explore whether the reported relationship holds at higher masses. In Fig. \ref{fig:dens}, the individual stellar mass density profiles for the HFF are presented (blue circles) along with the fits to the ICL component (red dashed line) and Fig. \ref{fig:fig6} shows the slope of the stellar mass density profiles of the ICL component as a function of $M_{200}$. The squares represent the IllustrisTNG data from \citet[][]{Pillepich2017b}, for a volume of $\sim100^3$ Mpc$^3$ (TNG100, black squares) and $\sim300^3$ Mpc$^3$ (TNG300, grey squares).

Note that the PSF in the z-band might have an effect in the slope of the stellar mass density, as the PSF sends light from the inner parts of sources to their outer parts modifying their intrinsic profile. However, the  PSF effect should not play a major role due to the smooth profile shape of the ICL (see a detailed discussion about this on the Appendix in Paper I).

\section{Discussion and Conclusions}

The results presented in this work show the extraordinary power of the HFF survey to address the origin and evolution of the ICL. This impressive data makes possible to explore the properties of the ICL in 6 very massive clusters in the interesting redshift range $0.3-0.6$; a period of time crucial to understand the formation of this elusive component in galaxy clusters. 

\subsection{Ages and metallicities}

Recently, numerical simulations have suggested that the origin of the ICL is the result of the disruption and tidal stripping of massive ($10^{10-11}$ $M_{\odot}$; stellar mass) satellite galaxies infaling in the cluster \citep[see also][]{Purcell2007, Martel2012, Contini2014, Cooper2015}. Since most of the ICL is produced by tidal stripping of massive satellites, this component is expected to have a metallicity that is similar to that of these galaxies \citep[][]{Contini2014}. That also naturally predicts a gradient in the ICL as the more massive galaxies, more metal-rich, are closer to the centre of the cluster \citep[by mass segregation, e.g.,][]{Presotto2012, Roberts2015}. The results presented in this paper are in good agreement with these predictions. Let's expand on this.

In Section \ref{agemets}, we present the age and metallicity profiles for the HFF clusters. We find negative age and metallicity gradients with radius for all the HFF clusters. On average, we are probing the ICL down to radial distances of $\sim 120$ kpc in all the clusters ($50<R<120$ kpc, orange area in Fig. \ref{fig:fig3}). The metallicity of the ICL ranges between [Fe/H]$_{ICL}$ = $-0.3$ to $-0.5$ (Z $\sim0.3-0.4$ Z$_\odot$). 
These metallicities are similar to those found in the outskirts of the Milky Way \citep[e.g.,][]{Cheng2012}, suggesting that the tidal stripping of MW-like galaxies is the main responsible for the formation of the ICL of these clusters. As the galaxies fall into the cluster potential, their less bounded material, their outskirts, will be stripped more easily. We have also found a gradient in metallicity from the center to the ICL region. This is a natural consequence of dynamical friction: the most massive galaxies approach the inner cluster regions faster than their less massive counterparts.

The mass-metallicity relationship of \citet[][]{Gallazzi2005} can be used as a proxy to derive the metallicity of the progenitors of the ICL. A galaxy with the metallicity of the ICL would have a typical stellar mass of $0.5-1.5\times 10^{10}$ $M_{\odot}$. This range of masses are slightly smaller but compatible with that of the Milky Way \citep[e.g., $6.4\times10^{10}$ $M_{\odot}$;][]{McMillan2011}. If we take into account that it is the outer regions of the galaxies the places that are more easily stripped then, taking into account the metallicity gradient of the galaxies, objects even more massive than $\sim$10$^{10}$ $M_{\odot}$ are good candidates to be the primordial source of metallicity for the ICL.

The observed ages of the stellar population of the ICL are between $2$ to $6$ Gyr younger than the BCG(s). This could be understood if active star forming galaxies orbiting the cluster ceased forming stars due to ram pressure stripping of their gas content \citep[e.g.][]{Boselli2009, Chung2009}. Then, those stars are stripped from the galaxies and become ICL. If we assume that the stellar populations of the BCG(s) were formed at $z\sim 2-3$, then the observed age difference  is compatible with the ICL being assembled at $z < 1$.
The measured ages for the ICL are consistent with the ages ($\sim2.5$ Gyr) derived spectroscopically by \citet[][]{Toledo2011} for a cluster at $z\sim0.29$  and \citet[][]{Adami2016} for a cluster at $z\sim0.53$ ($2.3$ Gyr). 

The metallicities derived for the ICL of the HFF are in agreement with previous observational studies for both nearby and similar redshift clusters (e.g., \citealt{Williams2007, Coccato2011}; Paper I; \citealt{DeMaio2015}). Recently, \citet[][]{Morishita2016} explored the colours of the ICL of the $6$ HFF clusters. They also found that only metallicity gradients are not able to reproduce the observed colours, but negative age gradients are also required. Their results suggest that the ICL is dominated by stars of ages $\sim 1-3$ Gyr, similar to the ages we found here for the ICL. Their inferred metallicities also agree with a subsolar metallicity.

The above numbers for age and metallicity are, of course, average values and do not describe entirely the large diversity of ages and metallicities expected in the ICL. In fact, both spectroscopic and red giant branch estimates of a handful of clusters have shown that the ICL corresponds to a mixture of young, intermediate and old, metal-poor and metal-rich stars, in varying fractions \citep[][]{Melnick2012, Edwards2016}. Because of the methodology we are using here, our estimates are luminosity-weighted (while in \citealt[][]{Melnick2012, Edwards2016} are mass-weighted). In this sense, our results should be understood as a description of the average properties of the ICL.

\subsection{Fraction of light in the ICL}

In Section \ref{fractions}, we derived the fraction of ICL to the total light of the cluster using several definitions. The most widely used definition from the observational point of view is to apply a cut in surface brightness and assume as ICL the light fainter than that threshold. Using this definition, we found that there is a slight increase in the fraction of ICL with cosmic time, i.e. decreasing redshift. Note, however, that we are exploring a narrow redshift range $0.3<z<0.55$ and we have only $6$ very massive clusters, therefore any conclusion must be taken with caution. This observed increase is in qualitative agreement with the results by \citet[][]{Krick2007} and \citet[][]{Burke2015}, who found that the ICL grows a factor of $\sim4-5$ between $0.2<z<0.4$.

There is a discrepancy between the values of the fraction of ICL for M0416 and AS1063 reported by \citet[][]{Burke2015} ($2.5\pm0.1$ and $6.4\pm0.1$) and the values we found here ($1.57\pm0.03$ and $3.24\pm0.06$). This disagreement is caused by the difference in magnitude threshold at defining the ICL, i.e. their threshold is brighter (they are including more light from the inner parts of the cluster). In addition, our masking criteria might have some effect. For example, because HFF images are  deeper than the ones in the CLASH survey, we needed to mask part of M0416 cluster to avoid contamination of a bright star in the FOV (see Section \ref{clusters}).

In the case of our lowest redshift cluster, A2744, the HFF survey only observed one of the substructures. Using a larger FOV, including most of the cluster, \citet[][]{Krick2007} found a fraction of $14\%$ in the r-band and $11\%$ in the B-band for this cluster. This is in tension with our measurement ($<$5\%) for this subregion of A2744. The reason of this discrepancy is, again, the different surface brightness cuts. Their surface brightness threshold is $25.7$ mag/arcsec$^2$ in the B-band, which translates in $\sim$25 mag/arcsec$^2$ in the V-band, assuming B-V = $0.7$ ($2$ Gyr and [Fe/H] =$-0.4$, \citealt{Vazdekis2016}). This is one magnitude brighter than in our case. Therefore, they are including more light from the central parts of the cluster than us. On the other hand, \citet[][]{Jimenez-Teja2016} explored the ICL fraction of A2744 using the images from the HFF survey. They used Chebyshev-Fourier functions to disentangle between BCG and ICL without prior assumptions on the properties of the system. They found an ICL fraction of $19.2\pm2.9\%$. In this case, the main difference with our result comes from the inclusion of ICL light that is embedded (in projection) into the BCG and a simple surface brightness threshold is unable to account for.

The methodology used in \citet[][]{Jimenez-Teja2016} is conceptually similar to the two-component profiles used in previous ICL studies \citep[e.g.][]{Gonzalez2005, Zibetti2005, Giallongo2014}. To explore how this measurement of the ICL could affect our results, we fitted a linear profile to the surface brightness distribution in the V-band of each of the clusters at $R>50$ kpc, see Section \ref{fractions}. On doing this, we are able to account for the light of the ICL which is (in projection) in the central regions of the cluster but is outshined by the brightness of the brightest galaxies. Once this missing ICL light is included, we found that on average, the fraction of ICL is $\sim7\%$ (for $R<R_{500}$). 
\citet[][]{Morishita2016} also explored the fraction of ICL of the HFF clusters. Their methodology also accounts for the light embedded in the BCG. Their ICL fractions for $R<500$ kpc compared to our $R<R_{500}$ estimations are in agreement with the exception of two clusters: M0717 and A370. For A370, one possible explanation for the discrepancy could be the different choice of BCGs (two in our case) that might be responsible for the lower fraction of ICL in our measurements. For M0717, our more conservative masking criteria could be causing the disagreement.

Observations and simulations \citep[e.g.][]{Seigar2007, Donzelli2011, Pillepich2014, Cooper2015, Pillepich2017b} suggest a two-component description of the BCG+ICL light profile, with the outer component exhibiting an exponential form in most cases. Note, however, that to assume an exponential profile for the ICL in the inner regions is something that observations alone can not prove. Consequently, simulations are needed to see whether assuming an exponential form for the entire light profile of the ICL is justified or not. 

Using an exponential fit to describe the ICL, we find only a mild evolution of the fraction of light in the ICL with cosmic time. That is in agreement with the lack of evolution in the fraction observed by \citet[][]{Guennou2012} between $z $=$ 0$ and $z $=$ 0.8$. However, a redshift dependence, such as the one observed here when using the interval $26<\mu_V<27$ mag/arcsec$^2$ for the defining the ICL and in \citet[][]{Burke2015}, might be caused (at least partially) by a bias connected with the way the ICL is measured using a surface brightness cut. At higher redshift, stellar populations are younger (both in the ICL and in the BCGs) and therefore brighter in the optical bands while at lower redshift, ages get older (particularly for the BCGs) and the stars fainter at optical wavelengths. If a fixed surface brightness limit is used for defining the ICL  then, the location (in radial position) of the isophote of a given surface brightness will be closer to the centre as comic time progresses. This is due to the brightness of the BCGs gets dimmer with cosmic time much faster than the ICL (which is continuously forming by the accretion of new, younger, satellites). This effectively includes more light as ICL at lower redshifts. This effect is more pronounced in the optical bands than in the IR, especially in the B-band. That could explain the steep increase in the ICL fraction observed in \citet[][]{Burke2015} (B-band) compared to the increase observed in this work (V-band). 

Although, an increase in the fraction of ICL with time is expected from simulations \citep[e.g.][]{Rudick2011, Contini2014}, we should be careful when comparing our results with the simulations. Ideally, one would like to measure the ICL light fractions in the rest-frame IR bands, to reduce the effect of the evolution of age that affects drastically the brightness of the stellar populations in the optical. Alternatively, one could try estimating the ICL stellar mass fraction. Finally, accounting for the missing ICL light projected in the BCG might also diminish this effect.

Relaxation may also play a role in the fraction of ICL in a cluster. Relaxed clusters are dynamically older clusters that have already been through significant mergers. Therefore, one could expect that they would have on average higher ICL fractions \citep[][]{Rudick2011, Cui2014}. There is a slight evidence that the most relaxed cluster in our sample, AS1063, has a higher fraction of ICL compared to the other clusters. Nonetheless, with a sample of only $6$ clusters is still premature to conclude whether relaxation plays a role or not. In the future, larger samples of clusters, with different dynamical states, will clarify this issue. Finally, we want to remind the reader that our conclusions are based on ages and metallicities that are derived using an SSP, i.e. they work as average values of the population. Naturally, this is an oversimplification. A full understanding of the stellar population properties of the ICL  requires the use of spectroscopy.

\subsection{The slope of stellar mass density of the ICL}

In Section \ref{slopes} and inspired by the results of \citet[][]{Pillepich2014, Pillepich2017b}, we explored the relationship between the slope of the stellar mass density profile of the ICL, $\alpha_{3D}$, and the total mass of the system. In Fig \ref{fig:fig6}, we see that the HFF clusters follow the extrapolation of the theoretical expectation between the $\alpha_{3D}$ and $M_{200}$. The range of slopes we measured in the HFF clusters is $-3.5<\alpha_{3D}<-2.5$. That means that the stellar halo of the clusters could be as shallow as the underlying dark matter halo, i.e. they could have similar slopes (the dark matter slopes range between $-2.6$ to $-2$, \citealt[][]{Pillepich2014}). That both components have similar shapes can be explained because more massive halos tend to accrete more and more luminous satellites \citep[e.g.][]{Gao2004} at recent times. Those satellites tend to deposit their stars at large radii ($\sim 100$ kpc, \citealt[][]{Cooper2015}) forming a less centrally concentrated stellar profile. The ages and metallicities reported for the ICL also agree with this scenario.

In this paper, we have studied the ICL of the $6$ Hubble Frontier Fields clusters. Taking advantage of their exquisite depth and multiwavelength coverage we have explored the properties of the stellar populations of this diffuse component with an unprecedented accuracy. We find that:
\begin{itemize}
\item The average metallicity of the ICL is [Fe/H]$_{ICL} \sim$ -0.5. This value is similar to that measured in the outskirts of the MW, suggesting that material stripped from MW-like satellites can be the dominant component of the ICL.
\item The average age of the stellar populations of the ICL is between 2 to 6 Gyr younger that the age of the central part of the BCG(s). Assuming that the ICL is not forming new stars, that suggests that the ICL is assembled at $z<1$.

\item Our results are compatible with no substantial increase in the fraction of light in the ICL with decreasing redshift. However, the redshift range explored is narrow ($<$2 Gyr) and the number of clusters limited to make a strong conclusion. 

\item To measure the evolution of the ICL light fraction with cosmic time, we discourage the use of blue bands. The blue filters are very sensitive to the evolution of the stellar populations, and so, imposing a surface brightness threshold to define the location of the ICL might introduce an artificial redshift dependence.

\item As predicted by simulations, the slope of the stellar mass density profile at high halo masses ($\sim10^{15}$) resembles the underlying dark matter profile, an indication of the origin of the ICL as debris of accreted satellites at recent times.

\end{itemize}

The results presented in this work show the extraordinary power of deep and multiwavelength surveys to address the origin and evolution of the ICL and the clusters themselves. In the future, larger samples will provide a more complete picture on the evolution of the ICL at different cluster halo masses and different redshifts.

\section*{Acknowledgements}

We thank the referee for constructive comments that helped to improve the original manuscript. We would like to thank STScI directors M. Mountain, K. Sembach and J. Lotz, and all the HFF team for making these extraordinary data available. We also thank Chris Mihos for sharing their simulation predictions and Javier Rom\'an for useful comments that helped us polish this manuscript. Support for this work was provided by NASA through grant HST-AR-14304 from the Space Telescope Science Institute, operated by AURA, Inc. under NASA contract NAS 5-26555 and by the Spanish Ministerio de Econom\'ia y Competitividad (MINECO; grant AYA2013-48226-C3-1-P). 
\\



\bibliographystyle{mnras}
\bibliography{ff_mnras.bib} 


\appendix

\section{Details on data reduction}

\subsection{PSF matching} \label{psf}

\begin{figure}
 \includegraphics[scale=0.105]{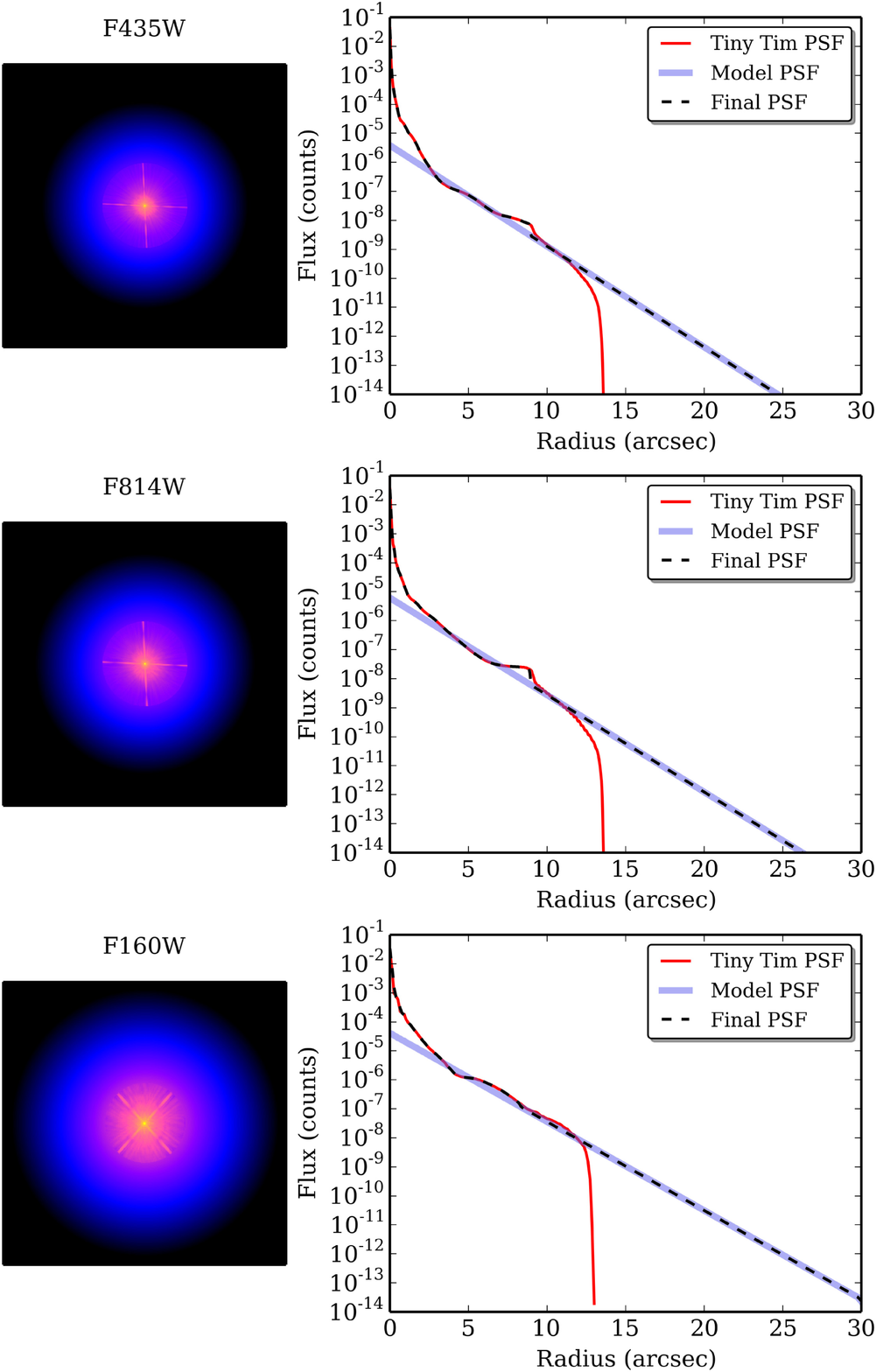}
 \caption{Extended PSFs for three different bands: F435W, F814W and F160W.  The left panels show the images of the PSFs while on the right panels, we plotted the radial profiles for the PSFs. The red solid line is the original Tiny Tim PSF that only extends $\sim 20 \times 20$ arcsec$^2$. The blue line is the exponential profile used as an extension for the PSF reaching the $60\times60$ arcsec$^2$. As seen, the extension reproduces the shape of the wings of the $\mathtt{TinyTim}$ PSF between $4$ to $12$ arcsec. Finally, the black dashed line is the radial profile of the combination of the $\mathtt{TinyTim}$ PSF until $15$ arcsec and the exponential profile down to $30$ arcsec in radius.}
 \label{fig:psf}
\end{figure}

The scatter light from nearby bright source affects dramatically low surface brightness features on the images, such as the ICL, as the contamination of the wings of the PSFs dominate at fainter magnitudes \citep[][]{deJong2008, Slater2009, Sandin2014}. It has also been shown that the wings of the PSF change with wavelength being more prominent at redder bands. To consistently derive age and metallicity profiles for the six HFF clusters, we followed this approach: we PSF-matched the images to the F160W image, which has the worst spatial resolution. By doing this, we ensure that our results are not artificially biased to redder colours due to the increasing prominence of the wings at longer wavelengths. \\
Obtaining a good PSF from the HFF fields is difficult due to the small field of view of the images (i.e. we have few bright stars to select) and the contamination of the wings of the PSF by the same ICL. Consequently, we used, for modeling the internal region of the PSF, a model PSF generated by the software $\mathtt{TinyTim}$. The size of the model PSFs was $\sim20\times 20$ arcsec$^2$.
The PSF derived from $\mathtt{TinyTim}$ was then rebinned to the current pixel size of our images ($0\farcs06$) and rotated according to the orientation of the camera. 
To derive realistic light profiles from the center of the galaxies down to the ICL in each band, it is important to accurately characterize the PSF of each image to large radial distances, larger than the size of the object to measure \citep[][]{Sandin2014}. As the effect of the wings is crucial for the goals of this work, we decided to extend the $\mathtt{Tiny Tim}$ PSFs using an exponential profile for both the ACS and the WFC3 PSFs. The choice of the exponential profiles was based on how well it reproduces the outer parts of the Tiny Tim PSF.  An example of the radial profiles for the final PSFs can be seen in Figure \ref{fig:psf}. The size of the final PSFs is $60\times 60$ arcsec$^2$.

\begin{figure}
 \includegraphics[width = 0.5\textwidth]{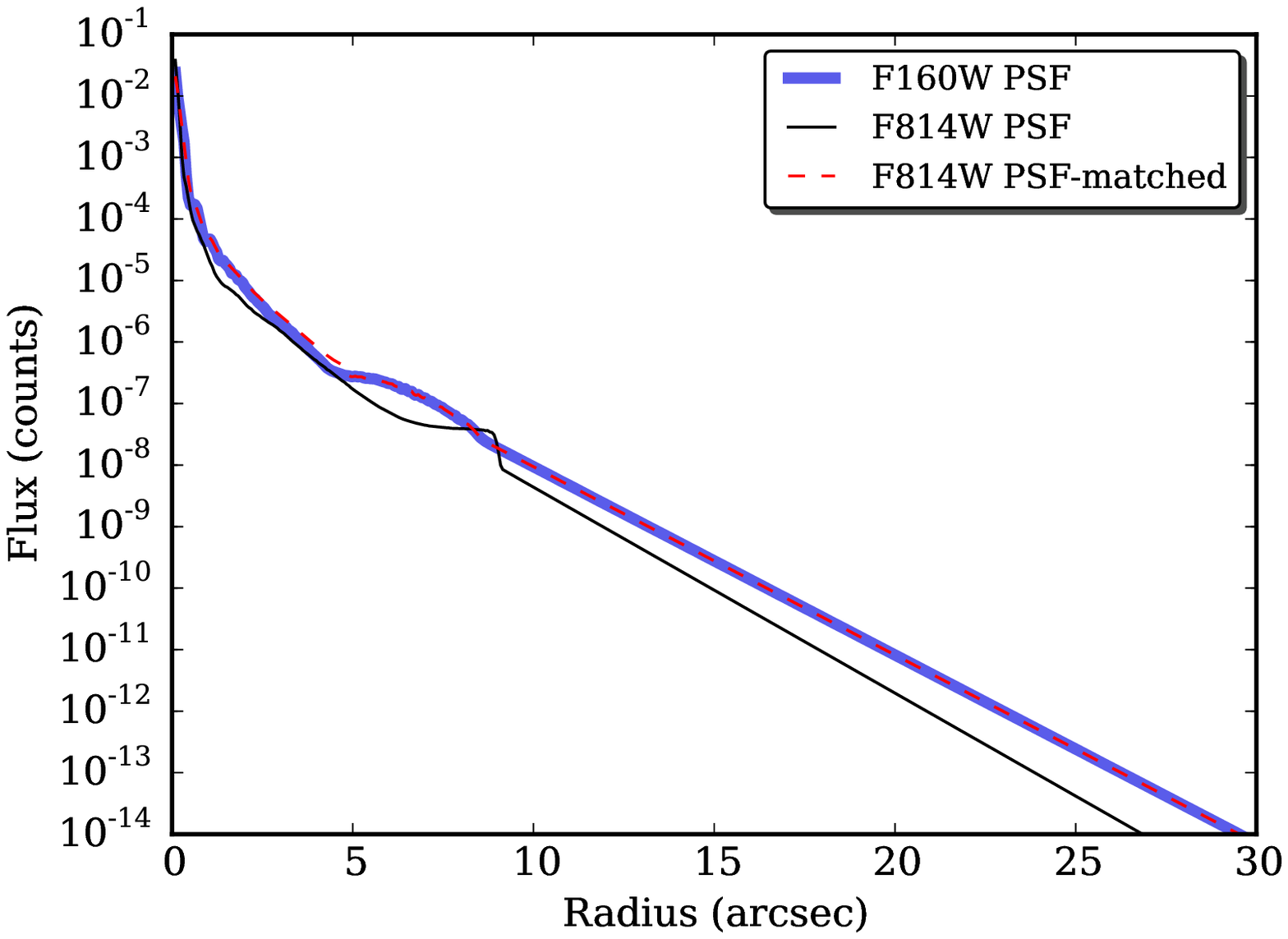}
 \caption{Comparison of the radial profiles of the convolved F814W PSF (dashed red line) with the F160W PSF (blue line). The original F814W PSF  is also plotted (black solid line).}
 \label{fig:psf_convolved}
\end{figure}

Once the extended PSF models were generated, we derived the kernels for each of the images using the IRAF task $\mathtt{lucy}$. Finally, once convolved with these kernels, the final images match the resolution of the F160W image. To check that the procedure was working correctly, we applied the previous kernels to our PSFs in each band. As expected, the convolution of the kernels with the PSFs produces the F160W PSF (see Fig. \ref{fig:psf_convolved}).

\subsection{Masking} \label{masks}

\begin{figure*}
 \begin{center}
  \includegraphics[scale=0.9]{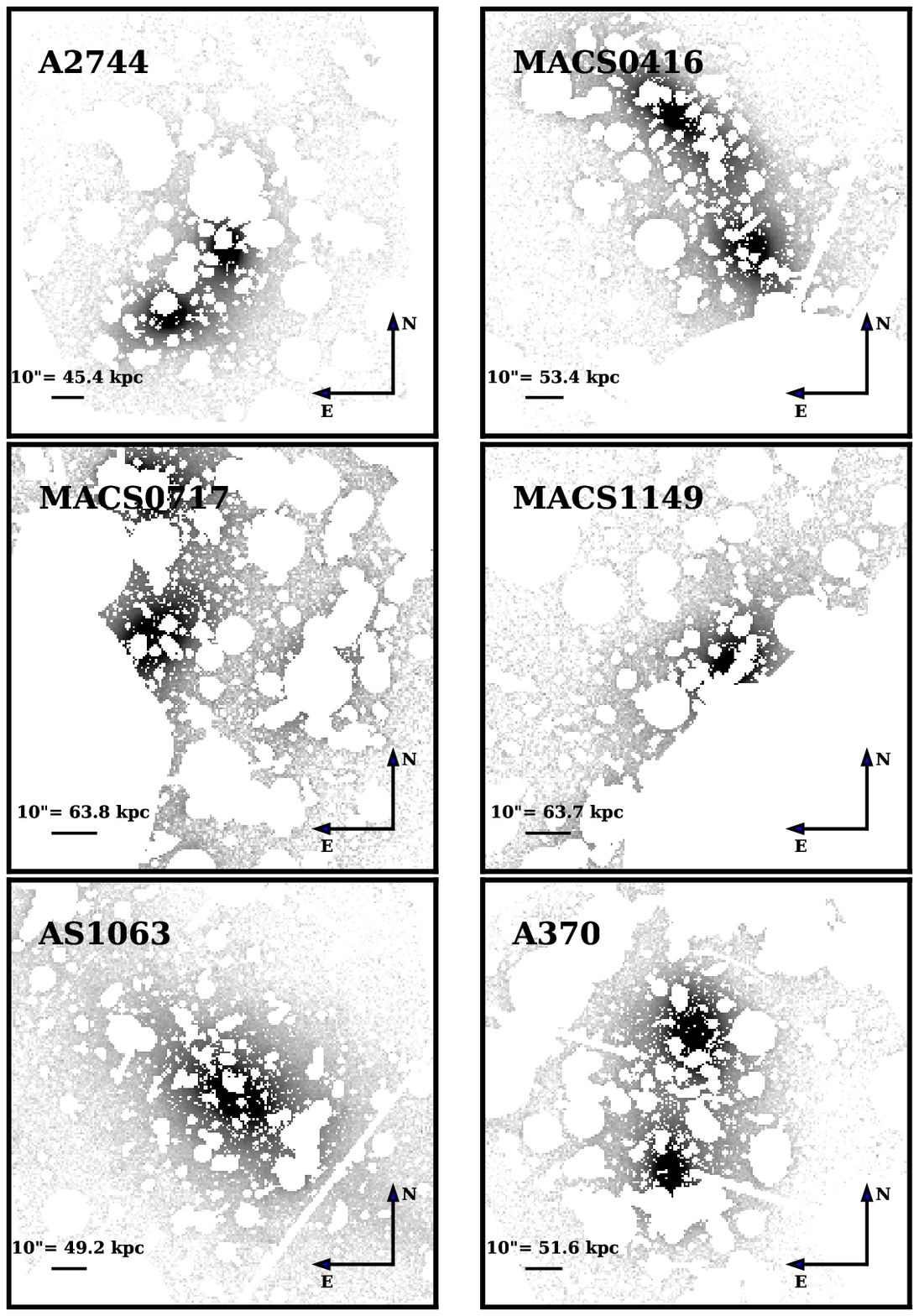}
  \caption{F160W masked images of the $6$ HFF clusters. In all the images, north is up and east is left. Images are $600$ kpc on each side. For M0416, M0717 and M1149, the presence of a bright foreground source close to the cluster made us mask a part of the image to avoid contamination of the ICL. } 
 \label{app:masks}
 \end{center}
\end{figure*}

In the case of deep surveys such as the HFF which aim to discover high redshift galaxies, the detection of sources must be optimized not only for those faint and small galaxies but also for large and closer objects. For that reason, we run SExtractor in the F160W image in a "hot+cold" mode, i.e. two separate SExtractor runs. The "cold" mode will detect the extended bright galaxies from the cluster while the "hot" mode is optimized to detect the faint and small sources. 
In our case, as the images are filled by the ICL, we run the "hot" mode in an unsharp masked image \cite[][]{Sofue1993}, to enhance the image contrast especially at the central parts of the clusters. To create the unsharp masked image, a gaussian filter with $\sigma$ = $3$ pix was convolved with the image and then subtracted from the original image. The "cold" mask was further expanded (dilated) $11$ pixels while the "hot" was dilated $2$ pixels. We, then combined the two masks to create the final mask for our images. Stars were masked manually, masking beyond their observed size to ensure little contamination from their wings. As the spikes of very bright stars can reach very far away, we masked around $\sim10$ arcsec from their cores to ensure that the light of the stars are no longer dominant. The extended spikes were masked separately as well. Other residual and foreground sources were also masked manually. In M0717 and M1149, very bright sources are located close to the centre of the clusters, i.e.: we have foreground galaxies and stars. We conservatively decided to mask part of the image, to ensure that no contamination will affect the measurement of the ICL in these clusters. The final masks were visually inspected to manually mask any remaining light that was missed by the process described above. This was repeated several times for each of the clusters to ensure that we minimize the contamination from galaxy outskirts and foreground and background light sources. 
The masked images can be seen in Fig. \ref{app:masks}. 

\section{Tabulated age and metallicity radial profiles for the HFF clusters}\label{agemettab}
The age and metallicity radial profiles for the HFF clusters are listed in Table \ref{table:agemet}. The ages and metallicities are the median ages and metallicities of the fits to the $500$ jackknife realizations of the SEDs, using the \citet[][]{Vazdekis2016} SSP models. The errors are the median errors divided by the square root of the number of realizations. 


\begin{landscape}
\begin{table}
 \tabcolsep 2.5pt
 \centering
 \begin{tabular}{@{}lcccccccccccc@{}}
      & \multicolumn{2}{c}{Abell 2744}  &  \multicolumn{2}{c}{MACSJ0416.1-2403}  &  \multicolumn{2}{c}{MACSJ0717.5+3745}  &  \multicolumn{2}{c}{MACSJ1149.5+2223}  &  \multicolumn{2}{c}{Abell S1063}  &  \multicolumn{2}{c}{Abell 370} \\
	Bin (kpc)	       &  Age (Gyr)   &   [Fe/H]       &  Age (Gyr)   &   [Fe/H] &  Age (Gyr)   &   [Fe/H]       &  Age (Gyr)   &   [Fe/H]    &  Age (Gyr)   &   [Fe/H]       &  Age (Gyr)   &   [Fe/H]    \\ \hline
0 - 0.5     &  $6.5 \pm 0.9 $  & $0.22 \pm  0.03$  &  $9.2 \pm 1.0 $  &  $0.00 \pm  0.09$&  $3.5 \pm 0.1 $  &  $0.07 \pm  0.04$ &  $3.5 \pm 0.2 $  &  $0.03 \pm  0.04$ &  $7.0 \pm 0.7 $  &  $0.19 \pm  0.03$ &  $3.5 \pm 0.7 $  &  $0.07 \pm  0.05$ \\
0.5 - 1     &  $6.8 \pm 1.0 $  & $0.19 \pm  0.04$  &  $6.0 \pm 0.9 $  &  $0.22 \pm  0.05$&  $3.6 \pm 0.4 $  &  $0.11 \pm  0.04$ &  $3.5 \pm 0.2 $  &  $0.03 \pm  0.04$ &  $7.0 \pm 0.7 $  &  $0.21 \pm  0.02$ &  $4.0 \pm 1.0 $  &  $0.21 \pm  0.05$ \\
1 - 1.5     &  $7.1 \pm 1.0 $  & $0.18 \pm  0.04$  &  $7.0 \pm 0.9 $  &  $0.16 \pm  0.04$&  $3.8 \pm 0.8 $  &  $0.12 \pm  0.05$ &  $3.6 \pm 0.3 $  &  $0.13 \pm  0.04$ &  $7.0 \pm 0.6 $  &  $0.21 \pm  0.02$ &  $5.3 \pm 1.0 $  &  $0.22 \pm  0.05$ \\
1.5 - 2.1   &  $7.0 \pm 1.0 $  & $0.17 \pm  0.04$  &  $6.9 \pm 0.9 $  &  $0.16 \pm  0.04$&  $5.6 \pm 0.8 $  &  $0.00 \pm  0.07$ &  $3.5 \pm 0.3 $  &  $0.00 \pm  0.04$ &  $7.0 \pm 0.7 $  &  $0.19 \pm  0.03$  &  $4.2 \pm 0.9 $  &  $0.22 \pm  0.05$ \\
2.1 - 3.1   &  $7.0 \pm 1.0 $  & $0.13 \pm  0.05$  &  $6.9 \pm 0.9 $  &  $0.14 \pm  0.05$&  $5.6 \pm 0.8 $  &  $-0.06 \pm  0.08$&  $3.4 \pm 0.5 $  &  $-0.01 \pm  0.04$&  $7.0 \pm 0.7 $  &  $0.19 \pm  0.03$  &  $4.1 \pm 0.9 $  &  $0.22 \pm  0.05$ \\
3.1 - 4.5   &  $4.0 \pm 1.0 $  & $0.22 \pm  0.05$  &  $5.4 \pm 0.9 $  &  $0.22 \pm  0.05$&  $3.5 \pm 0.9 $  &  $0.00 \pm  0.08$ &  $3.5 \pm 0.8 $  &  $-0.02 \pm  0.04$&  $7.0 \pm 0.8 $  &  $0.18 \pm  0.04$  &  $4.2 \pm 0.9 $  &  $0.22 \pm  0.05$ \\
4.5 - 6.6   &  $3.9 \pm 1.0 $  & $0.14 \pm  0.06$  &  $4.0 \pm 0.9 $  &  $0.22 \pm  0.05$&  $3.5 \pm 0.9 $  &  $-0.06 \pm  0.09$&  $3.5 \pm 0.9 $  &  $0.02 \pm  0.04$ &  $6.9 \pm 0.9 $  &  $0.16 \pm  0.04$  &  $4.2 \pm 1.0 $  &  $0.22 \pm  0.05$ \\
6.6 - 9.7   &  $3.6 \pm 1.0 $  & $0.08 \pm  0.07$  &  $3.8 \pm 1.0 $  &  $0.22 \pm  0.06$&  $3.5 \pm 1.0 $  &  $-0.17 \pm  0.10$&  $2.9 \pm 0.9 $  &  $0.21 \pm  0.04$ &  $5.9 \pm 0.9 $  &  $0.22 \pm  0.04$  &  $5.0 \pm 1.0 $  &  $0.21 \pm  0.06$ \\
9.7 - 14.1  &  $3.6 \pm 0.3 $  & $0.00 \pm  0.05$  &  $4.8 \pm 0.4 $  &  $0.15 \pm  0.03$&  $1.6 \pm 0.4 $  &  $-0.12 \pm  0.09$&  $3.0 \pm 0.5 $  &  $-0.01 \pm  0.04$&  $5.8 \pm 0.9 $  &  $0.22 \pm  0.04$  &  $5.4 \pm 0.9 $  &  $0.21 \pm  0.06$ \\
14.1 - 20.6 &  $3.4 \pm 0.4 $  & $-0.06 \pm 0.07$  &  $4.8 \pm 0.5 $  &  $0.14 \pm 0.04$&   $2.8 \pm 0.5 $  &  $-0.23 \pm  0.10$&  $3.0 \pm 0.6 $  &  $-0.11 \pm  0.04$ & $5.8 \pm 0.9 $  &  $0.22 \pm  0.04 $ &  $5.6 \pm 1.2 $  &  $0.22 \pm  0.07$ \\
20.6 - 30.1 &  $3.3 \pm 0.4 $  & $-0.12 \pm 0.08$  &  $2.7 \pm 0.5 $  &  $-0.07 \pm 0.09$&  $2.8 \pm 0.7 $  &  $-0.44 \pm  0.12$&  $2.8 \pm 0.4 $  &  $0.07 \pm  0.04$ &  $9.6 \pm 0.7 $  &  $0.07 \pm  0.04$ &  $3.2 \pm 0.9 $  &  $-0.05 \pm  0.08$\\
30.1 - 44   &  $4.0 \pm 0.4 $  & $-0.21 \pm 0.09$  &  $2.2 \pm 0.5 $  &  $-0.15 \pm 0.12$&  $1.7 \pm 0.3 $  &  $-0.39 \pm  0.09$&  $2.3 \pm 0.3 $  &  $0.07 \pm  0.04$ &  $3.6 \pm 1.1 $  &  $-0.03 \pm  0.08$&  $2.2 \pm 0.6 $  &  $0.22 \pm  0.10$\\
44 - 64.2   &  $2.5 \pm 0.6 $  & $-0.28 \pm 0.07$  &  $1.4 \pm 0.5 $  &  $-0.19 \pm 0.14$&  $2.4 \pm 0.6 $  &  $-0.19 \pm  0.14$&  $2.1 \pm 0.4 $  &  $0.07 \pm  0.04$ &  $2.8 \pm 0.6 $  &  $-0.12 \pm  0.09$ &  $2.3 \pm 0.6 $  &  $-0.19 \pm  0.13$\\
64.2 - 93.9 &  $2.5 \pm 0.6 $  & $-0.29 \pm 0.07$  &  $1.4 \pm 0.5 $  &  $-0.69 \pm 0.19$&  $1.8 \pm 0.7 $  &  $-0.27 \pm  0.25$&  $2.3 \pm 0.4 $  &  $0.07 \pm  0.04$ &  $1.8 \pm 0.1 $  &  $-0.12 \pm  0.07$ &  $1.7 \pm 0.6 $  &  $-0.38 \pm  0.17$\\
93.9 - 137  &  $3.6 \pm 0.8 $  & $-0.21 \pm 0.09$  &  $\cdots$         &  $\cdots$         &  $2.1 \pm 0.7 $  &  $-0.77 \pm  0.41$&  $1.6 \pm 0.4 $  &  $0.07 \pm  0.04$ &  $1.6 \pm 0.4 $  &  $-0.48 \pm  0.10$&  $1.2 \pm 0.1 $  &  $-0.50 \pm  0.25$\\
137 - 200   &  $3.6 \pm 0.9 $  & $-0.29 \pm 0.10$  &  $\cdots$         &  $\cdots$         &  $\cdots$         &  $\cdots$          &  $\cdots$         &  $\cdots$          &  $\cdots$         &  $\cdots$ &  $\cdots$         &  $\cdots$ \\ \hline  
\end{tabular}\caption{Age and metallicity radial profiles of the $6$ HFF clusters. The age and metallicities are the median ages and metallicities coming from $500$ jackknife realizations of the photometry of the clusters. The final radial bin explored ($R_{limit}$) is defined as the farthest spatial bin with accurate ages and metallicities (i.e. those bins where the number of reliable filters is $> 4$)}\label{table:agemet}  
\end{table}
\end{landscape}

\section{Stellar mass density profiles}\label{dens}

\begin{figure*}
\begin{tabular}{@{}cc@{}}
\centering
\begin{subfigure}[t]{0.45\textwidth}
\includegraphics[width=\textwidth]{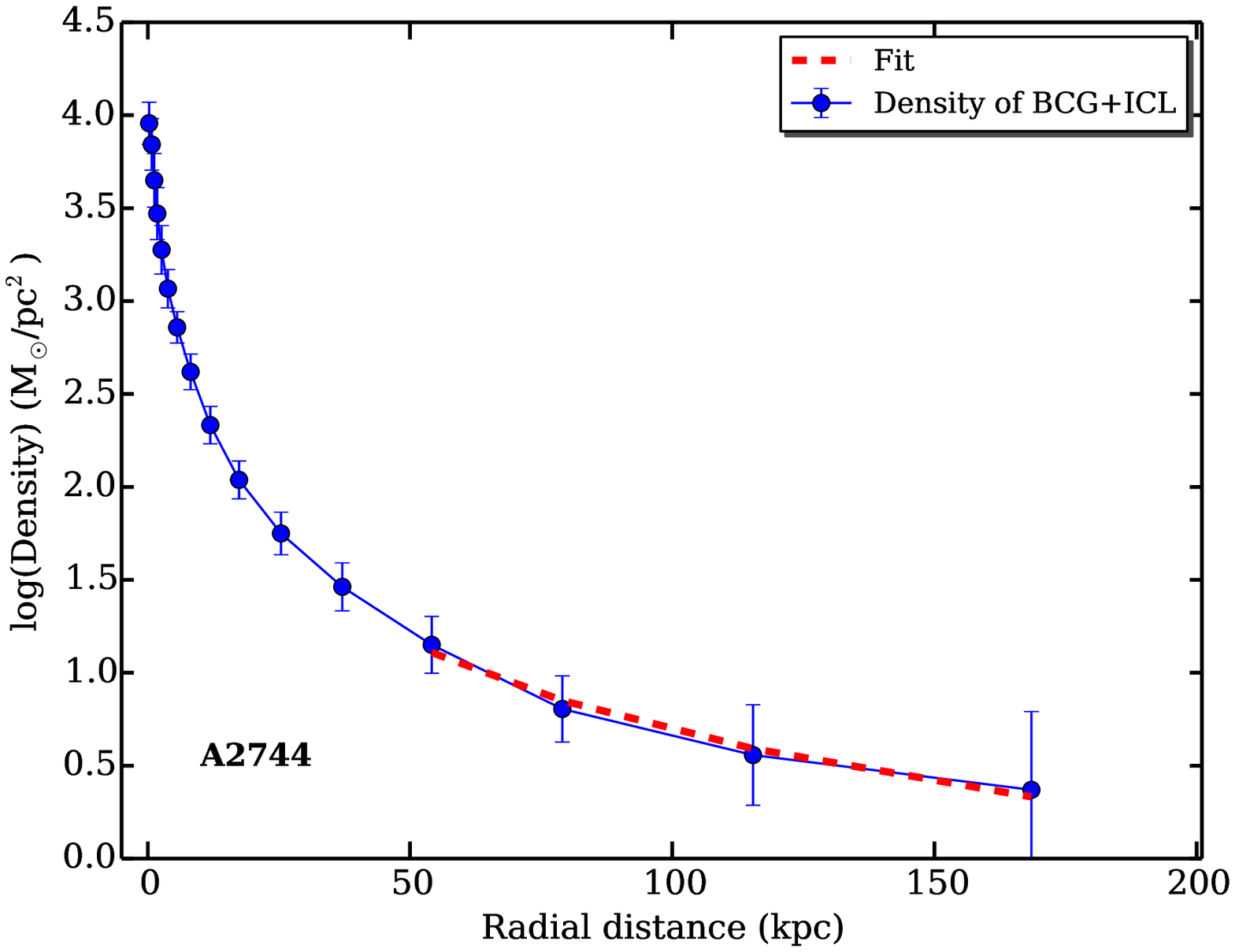} 
\end{subfigure}
&
\begin{subfigure}[t]{0.45\textwidth}
\includegraphics[width=\textwidth]{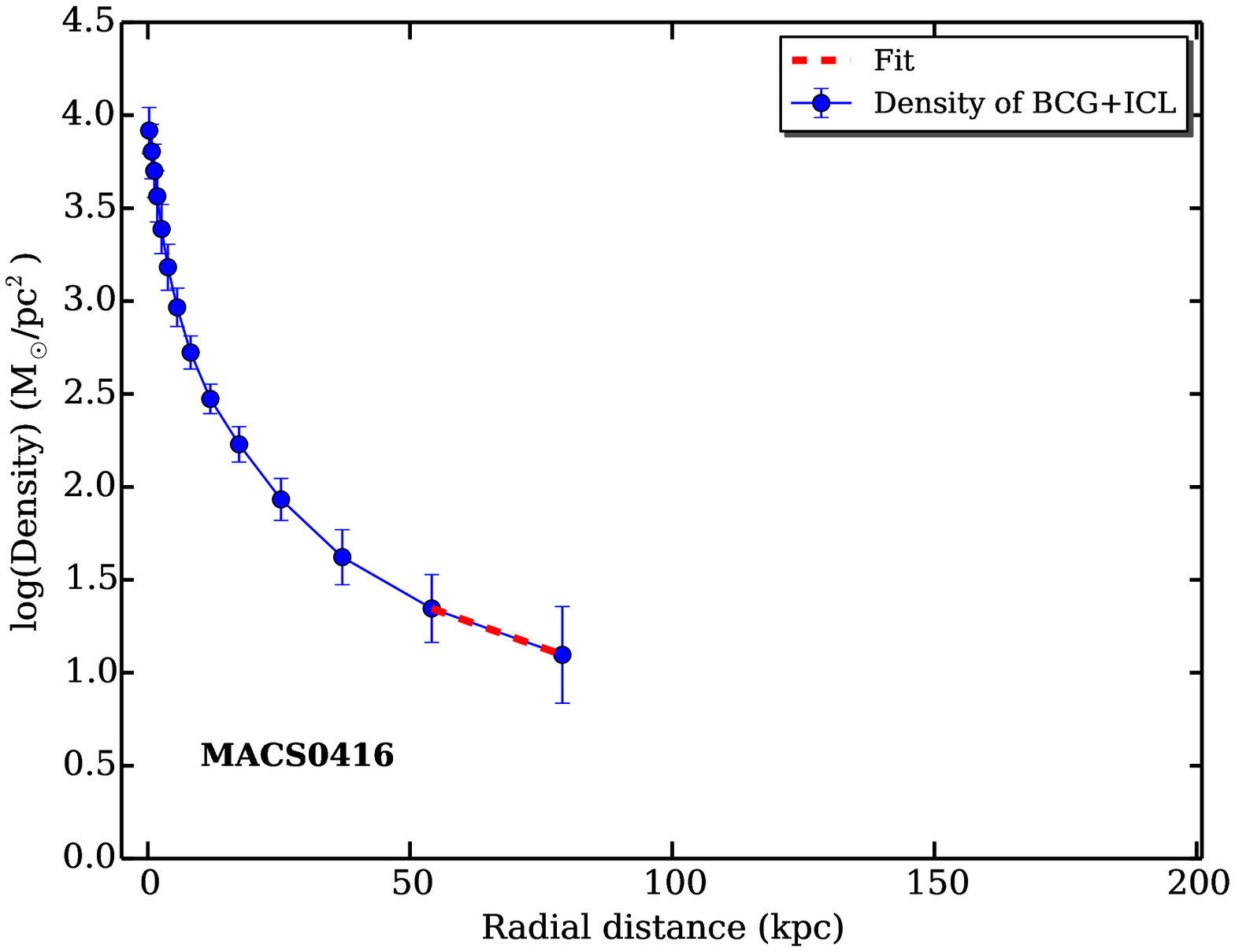} 
\end{subfigure}
\\
\begin{subfigure}[t]{0.45\textwidth}
\includegraphics[width=\textwidth]{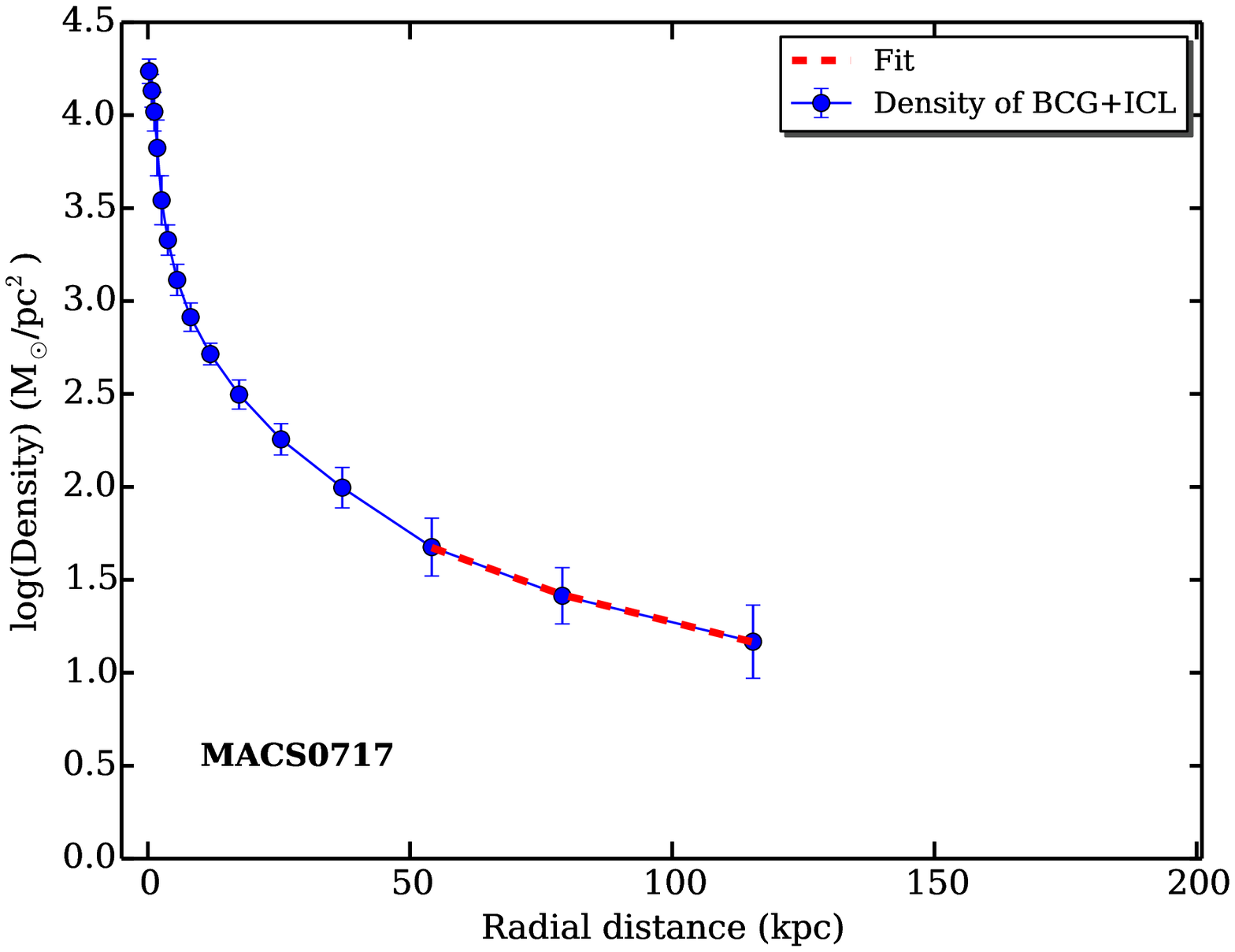} 
\end{subfigure}
&
\begin{subfigure}[t]{0.45\textwidth}
\includegraphics[width=\textwidth]{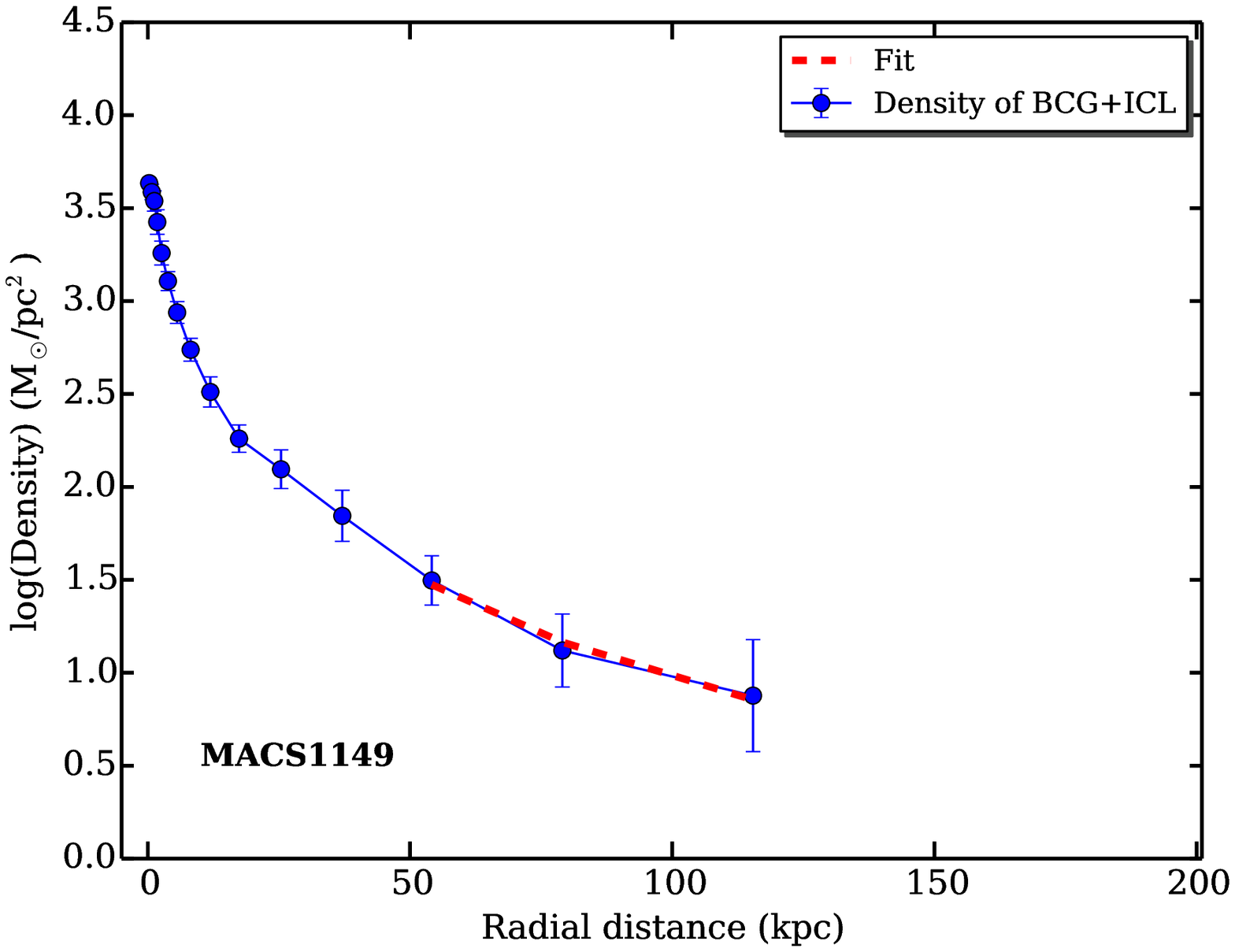} 
\end{subfigure}
\\
\begin{subfigure}[t]{0.45\textwidth}
\includegraphics[width=\textwidth]{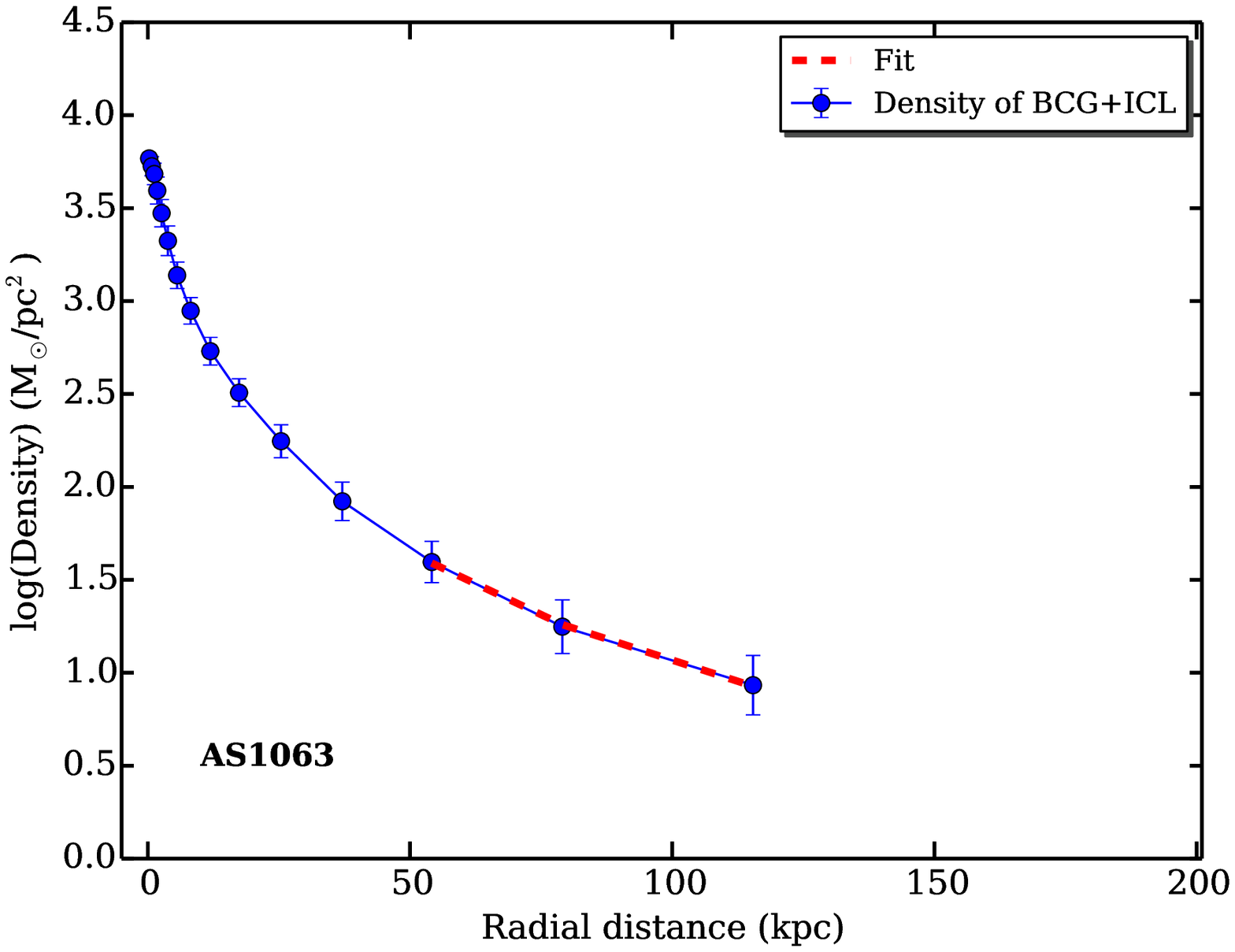} 
\end{subfigure}
&
\begin{subfigure}[t]{0.45\textwidth}
\includegraphics[width=\textwidth]{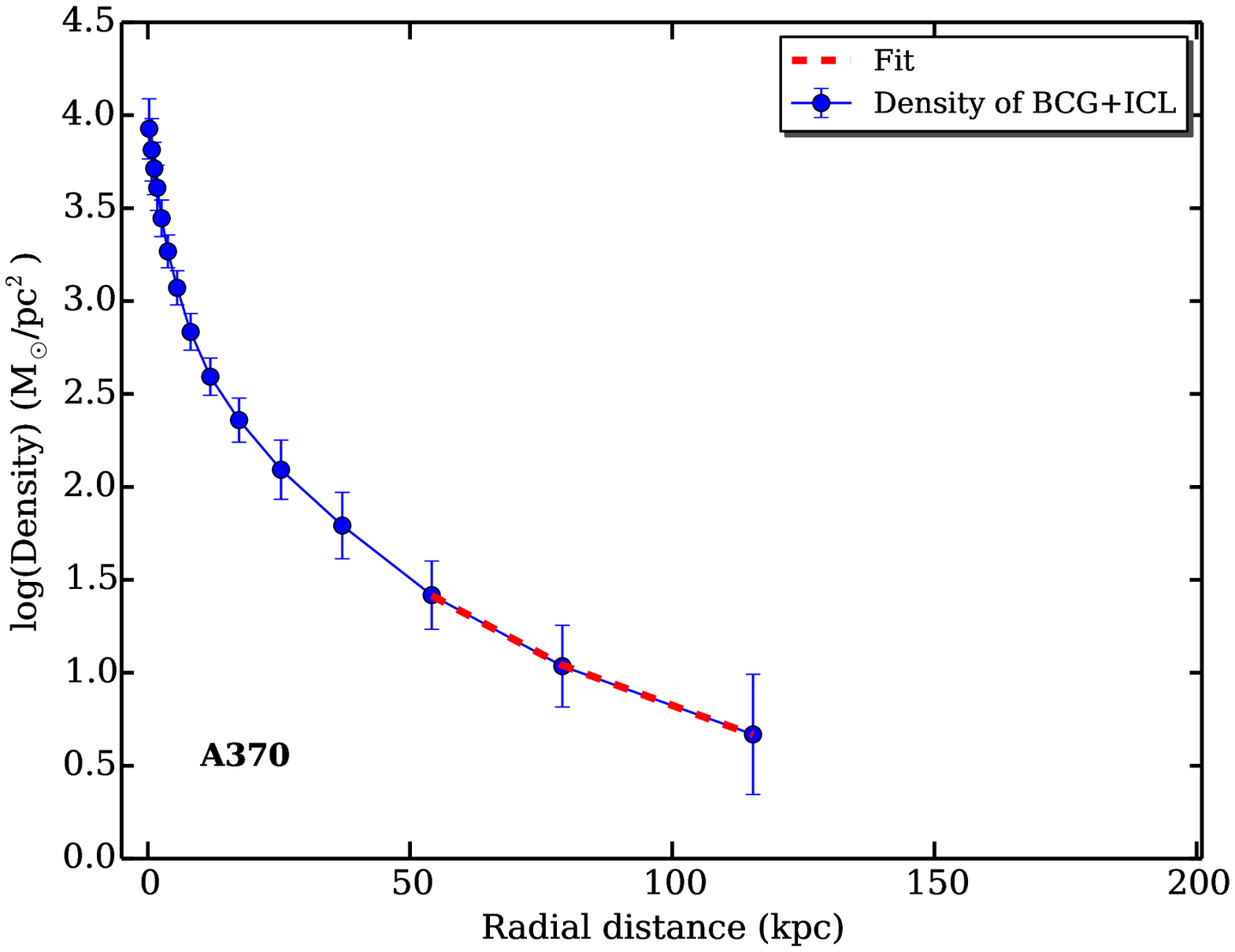} 
\end{subfigure}
\\
\end{tabular}  
 \caption{Stellar mass density profiles for the BCG(s) + ICL of the $6$ HFF clusters (blue points). The red dashed lines are the linear fits in log space to the ICL component $R>50$ kpc  from which we derived the slopes of the stellar mass density profiles for Fig. \ref{fig:fig6}.} 
 \label{fig:dens}

\end{figure*}

In Fig. \ref{fig:dens}, we present the stellar mass density profiles of each of the HFF clusters. The stellar mass density ($\rho$ in $M_{\odot}$/$pc^2$) was computed using the same approach as in Paper I but using the i-z colours to compute the M/L ratio (see Sec. \ref{slopes}).
The blue filled circles are the observed profile of BCG(s) + ICL while the red dashed line is the fit to the profile in the range $50$ kpc $<R<R_{limit}$.

\section{Total stellar mass of the cluster and completeness} \label{completeness}

In this section, we provide an estimate on the bias in the total stellar mass of our clusters caused by the different mass completeness due to their different redshifts. The measure of the total stellar mass depends on the redshift as we are missing more faint galaxies in our high redshift clusters compared to the closest ones. A detailed account for this effect is not trivial in our analysis, but we provide here with a rough estimation of how this can contribute in our analysis.

Our closest cluster, A2744 is at $z$=$0.308$, whereas the farthest cluster, M0717 is located at $z$=$0.545$. This means a difference in distance modulus between both clusters of DM=$42.48-41.02$=$1.46$. In this sense, at quantifying the contribution of light from detected galaxies in our closest cluster, we are including galaxies, on average, which are 1.46 mag fainter than the faintest in the farthest cluster. In terms of stellar mass (assuming a similar M/L), this means that we detect galaxies which are 4 times less massive in our closer cluster than in our farthest one.
The faintest galaxies we detect as members of our clusters has a typical magnitude of $r\sim28$ (restframe), which at the redshift of our closest cluster is equivalent to $M_r$=$28-41.02$=$-13.02$ and for the farthest cluster $M_r$=$28-42.49$=$-14.49$. Assuming that the stellar mass function of the galaxies in these clusters has not changed dramatically in such redshift interval and that they look similar to the stellar mass distribution of nearby rich clusters \citep[e.g. Fig. 2 in][]{Lan2016}, then we can have a rough estimation of the stellar mass we are missing in the interval $-14.5$ to $-13$ mag. Using Fig. 4 from \citet[][]{Lan2016}, and assuming the Conditional Luminosity Functions form holds for fainter magnitudes, we can see that for the most massive cluster ($\log M_{200}\sim14.9$, bottom right panel), the total number of galaxies should double from $-14.5$ to $-13$. In this sense, the missing mass located in less massive galaxies in our farthest clusters would be of the order of $50\%$ (i.e. $2\times \frac{1}{4}$) less. So, if anything, the fraction of ICL in our high-z clusters could be overestimated by $50\%$ compared to the value we would have got taking into account the missing galaxies.



\bsp	
\label{lastpage}
\end{document}